\documentclass[
superscriptaddress,
reprint, amsmath,amssymb,
pra,
floatfix,
 array,
]{revtex4-1}
\usepackage{graphicx}
\usepackage{color}
\usepackage{enumerate}
\usepackage{graphicx}
\usepackage{ascmac}

\usepackage{enumerate}
\usepackage{lastpage}
\usepackage{times}
\usepackage{amsmath}

\begin{document}

\title{State injection, lattice surgery and dense packing of the deformation-based surface code}
\author{Shota Nagayama}
\email{kurosagi@sfc.wide.ad.jp}
\affiliation{Graduate School of Media and Governance, Keio University, 5322 Endo, Fujisawa-shi, Kanagawa 252-0882, Japan}
\author{Takahiko Satoh}
\affiliation{Keio Research Institute at SFC, Keio University, 5322 Endo, Fujisawa-shi, Kanagawa 252-0882, Japan}
\author{Rodney Van Meter}
\affiliation{Faculty of Environment and Information Studies, Keio University, 5322 Endo, Fujisawa-shi, Kanagawa 252-0882, Japan} 
     \begin{abstract}
	Resource consumption of the conventional surface code is expensive,
	in part due to the need to separate the defects
	that create the logical qubit far apart on the physical qubit lattice.
	We propose that instantiating the deformation-based surface code using superstabilizers
	makes it possible to detect short error chains connecting the superstabilizers,
	allowing us to place logical qubits close together.
	Additionally, we demonstrate the process of conversion from the defect-based surface code, which works as arbitrary state injection,
	and a lattice surgery-like CNOT gate implementation that requires fewer physical qubits than the braiding CNOT gate.
	Finally we propose a placement design for the deformation-based surface code and analyze its resource consumption;
	large scale quantum computation requires
	$\frac{25}{4}d^2 +5d + 1$ physical qubits per logical qubit where $d$ is the code distance,
	whereas the planar code requires $16d^2 -16d + 4$ physical qubits per logical qubit,
	for a reduction of about 55\%.
     \end{abstract}
\maketitle
\vspace{3mm}

\section{Introduction}
The surface code is to date the most feasible proposal
~\cite{Fowler:2009High-threshold_universal_quantum_computation_on_the_surface_code,
kitaev2003ftq,bravyi1998qcl,raussendorf07:_2D_topo,raussendorf07:_topol_fault_toler_in_clust}
to tolerate
the inevitable imperfections in qubit states in a quantum computer
~\cite{nielsen-chuang:qci,VanMeter:2013:BBQ:2507771.2494568,cirac2000scalable,
yao10:_scal_room_temp_arch,chiaverini05:qft-impl,PhysRevB.76.174507,
stace:200501,Barrett:PhysRevLett.105.200502,
ladd10:_quantum_computers}.
The surface code has advantages for implementation over other quantum error correcting codes;
the surface code requires only a 2D lattice of physical qubits with nearest-neighbor interactions,
sustains scalability across a large range since the surface code can be enlarged by lengthening the columns and the rows of the 2D lattice,
and has higher error threshold than other codes.
There are several proposals for producing a logical qubit on the surface code lattice;
the planar code
~\cite{:/content/aip/journal/jmp/43/9/10.1063/1.1499754}
and the defect-based code
~\cite{raussendorf07:_2D_topo}
achieve universality by providing arbitrary state injection and
a basic set of one- and two-qubit
fault-tolerant gates.

Bombin and Delgado introduced another way to produce a qubit on the surface code, the deformed surface code,
and showed  Clifford gates and initialization to $|0\rangle$ and $|+\rangle$ 
~\cite{1751-8121-42-9-095302}.
They demonstrated a CNOT gate by braiding, which can be executed between
two logical qubits in the deformed surface code and
even between the deformation-based code and the defect-based code.
Since a SWAP gate can be implemented with three CNOT gates,
arbitrary state injection to the deformation-based code can be achieved
utilizing this heterogeneous CNOT gate.
First, use the standard state injection method in the
defect-based code, then swap into the deformation-based code.
However, this method is an indirect way to achieve state injection to the deformation-based code.

In this paper we show a conversion from the defect-based code to the deformation-based code
that enables the deformation-based code to hold an arbitrary state,
and demonstrate that a crossed pair of an $X$ superstabilizer and a $Z$ superstabilizer produces a deformation-based qubit,
without sacrificing the advantages above.
We employ the fault-tolerant stabilization utilizing a cat state generated by parallel $ZZ$ stabilization.
Additionally, we demonstrate a lattice surgery-like CNOT gate for the deformation-based code~\cite{Horsman:2012lattice_surgery}.
Lattice surgery is a non-transversal, scalable means of executing a CNOT gate on the planar code that
requires fewer resources than the ``braiding'' of the defect-based code.
Our lattice surgery-like CNOT gate for the deformation-based code requires fewer qubits than the conventional braiding.
Nevertheless, the error suppression ability is similar to conventional surface code
since the logical state is protected by normal stabilizers.
Our proposals may reduce the resource requirements of the surface code
in spatial accounting.

\section{Overview of the deformation-based surface code}
\label{sec:logical_qubit}
\begin{figure}[t]
 \begin{center}
  \includegraphics[width=8cm]{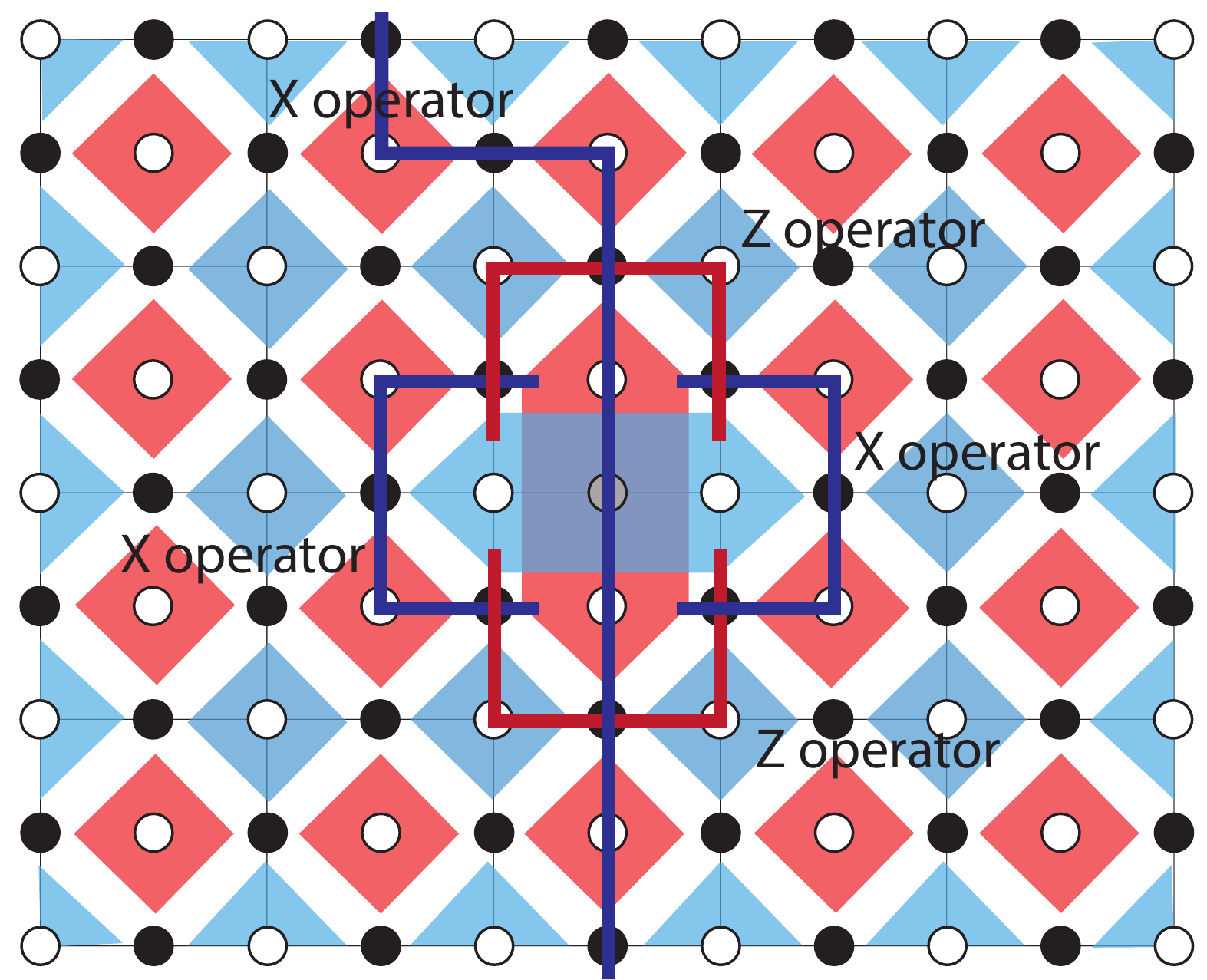}
  \caption{The deformation-based qubit of distance 3.
  Black dots depict data qubits and white dots are ancilla qubits.
  Each red diamond describes a $Z$ stabilizer and each blue diamond describes an $X$ stabilizer.
  The gray dot in the center depicts the unused data qubit, and
  the two 4-qubit $Z$ stabilizers the unused data qubit originally belonged to are merged to form the 6-qubit $Z$ stabilizer shown.
  The two 4-qubit $X$ stabilizers the unused data qubit originally belonged to are also merged to form the 6-qubit $X$ stabilizer shown.
  The thick lines are logical operators of the superstabilizer qubit.
   Any of the blue or the red paths serves as a logical $X$ operator
  or a logical $Z$ operator, respectively.
  } 
  \label{fig:integrated_qubit}
 \end{center}
\end{figure}

Figure \ref{fig:integrated_qubit} shows a distance 3 deformation-based qubit,
existing on the surface code lattice.
The surface code uses physical qubits placed on a 2D lattice.
The black dots are data qubits, and the white dots are ancilla qubits.
The lattice is separated into plaquettes as shown by black lines in the Figure.
Basically, each ancilla qubit is used to measure a stabilizer of the surrounding four data qubits.
A stabilizer $U$ is an operator which does not change a state,
\begin{equation}
 U\vert \psi \rangle = \vert \psi \rangle.
\end{equation}
An ancilla qubit in the center of a plaquette is used to measure the eigenvalue of a $Z$ stabilizer such as
$Z_aZ_bZ_cZ_d$ where $a\sim d$ denotes the surrounding four data qubits.
An ancilla qubit on the vertex is used for an $X$ stabilizer.

The number of logical qubits $k$ on a state of $n$ physical qubits is $k=n-s$ where $s$ is the number of independent stabilizers.
In Figure \ref{fig:integrated_qubit}, there are 48 data qubits, 19 $Z$ stabilizers
and 28 independent $X$ stabilizers, 
since any of the $X$ stabilizers is the product of all the others,
leaving a single degree of freedom for one logical qubit.

Two $Z_L$ operators of a deformation-based qubit
are shown in Figure \ref{fig:integrated_qubit}, either of which acts
on the logical qubit.
Three $X_L$ operators are shown in the figure, also working on the same logical qubit.
Two of the $X_L$ operators are the same shape as the described $Z_L$ operators,
while the third crosses the $Z$ superstabilizer ends at the boundaries of the lattice.
Those two $Z_L$ and two $X_L$ logical operators surrounding the superstabilizers correspond to the logical operators shown in
Figure 5 (a) in~\cite{1751-8121-42-9-095302}, except that our deformation-based qubit employs superstabilizers.
As with other surface code qubits, the products of a logical operator and stabilizers produce the redundancy for measurements
of logical operators.

The conventional implementation of a CNOT gate is separation and braiding.
See Ref.~\cite{1751-8121-42-9-095302} for details.

\begin{figure}[t]
 \begin{center}
  \includegraphics[width=8cm]{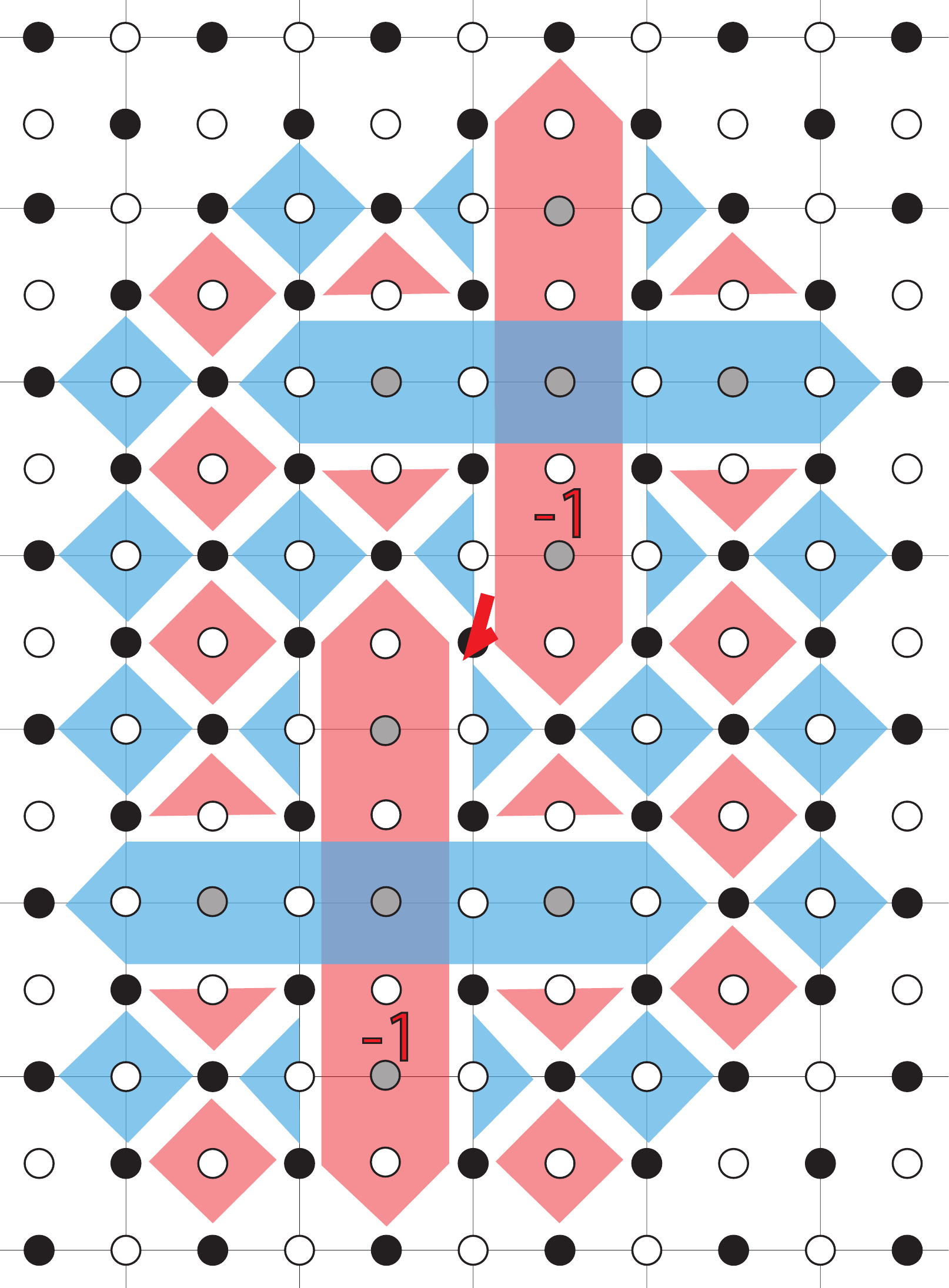}
  \caption{Neighboring distance 5 deformation-based qubits.
  Placement code distance apart from the boundary of the lattice is assumed.
  An $X$ error on the marked qubit results in $-1$ eigenvalues of the two red $Z$ superstabilizers.
  The two-defect surface code cannot correct an $X$ error on a data qubit which belongs to two defects,
  but the superstabilizers of the deformation-based code can.
  }
  \label{fig:correction}
 \end{center}
\end{figure}

Figure~\ref{fig:integrated_qubit} shows another important characteristic of the deformation-based qubit,
how to count its code distance.
Each logical operator consists of operations on three physical qubits, therefore the code distance of this deformation-based qubit is three.
An example of a longer code distance is shown in Figure~\ref{fig:correction},
which depicts two deformation-based qubits of distance five.

Figure~\ref{fig:correction} shows an advantage of deformation-based qubits compared to defect-based surface code qubits.
The deformation-based qubit exists at the junction of two superstabilizers,
so that every data  qubit alive in the lattice belongs to two $X$ stabilizers and two $Z$ stabilizers.
The two $Z$ superstabilizers find the $X$ error on the marked qubit in Figure~\ref{fig:correction},
hence the deformation-based qubits can be placed close to each other without being susceptible to logical errors,
though other surface code qubits must be placed code distance away.

\section{Transformation}
\begin{figure}[t]
 \begin{center}
  \includegraphics[width=8cm]{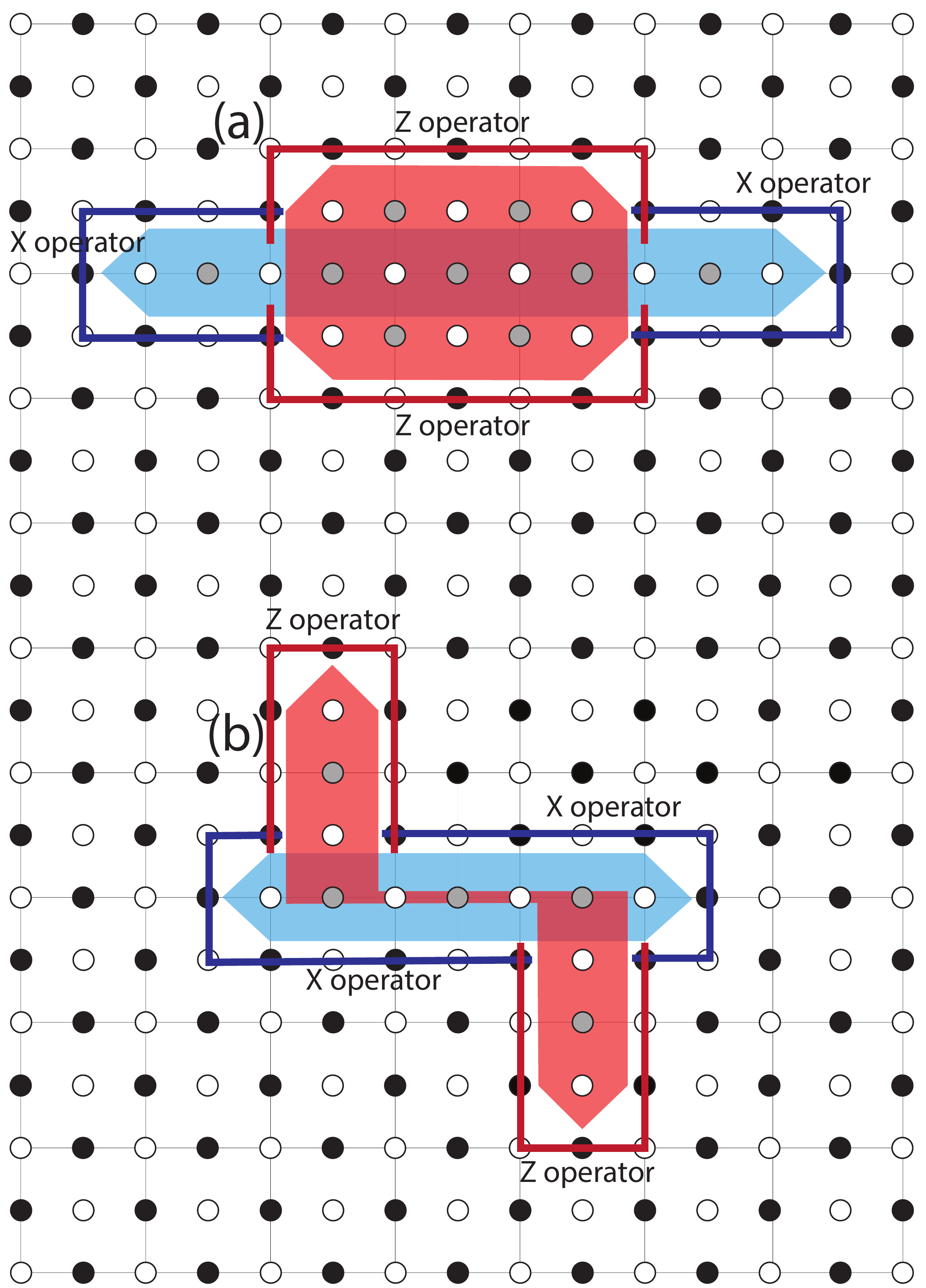}
  \caption{(a) ''Bar form'' deformation-based qubit, which has code distance 5.
  The $Z$ ($X$) logical operators exists between halves of the $X$ ($Z$) superstabilizer separated by the $Z$ ($X$) superstabilizer.
  (b) The deformation-based qubit of code distance 5, which has ``skew fin''.
  The $Z$ ($X$) logical operators exists between halves of the $X$ ($Z$) superstabilizer separated by the $Z$ ($X$) superstabilizer.
  }
  \label{fig:transform1} 
 \end{center}
\end{figure}

We have shown the ``four fin'' style deformation-based qubits.
Figure \ref{fig:transform1} shows two transformed deformation-based qubits of distance 5.
The deformation-based qubit in Figure \ref{fig:transform1} (a) is extended in the horizontal direction and
compressed in the vertical direction.
The perimeter of the $Z$ ($X$) superstabilizer can be considered to be separated by the $X$ ($Z$) superstabilizer.
The logical $Z$ ($X$) operator exists at any path connecting the separated halves.
The deformation-based qubit in Figure \ref{fig:transform1} (b) has a single, skewed $Z$ superstabilizer.
This transformation is achieved with more or less the defect-moving operations
of the defect-based surface code~\cite{Fowler:2009High-threshold_universal_quantum_computation_on_the_surface_code}.
The only difference is that the defect 
that does not have a stabilizer measurement
is replaced with the superstabilizer here.

\section{Conversion from a two-defect-based qubit}
\label{sec:state_injection}
\begin{figure}[t]
 \begin{center}
  \includegraphics[width=6cm]{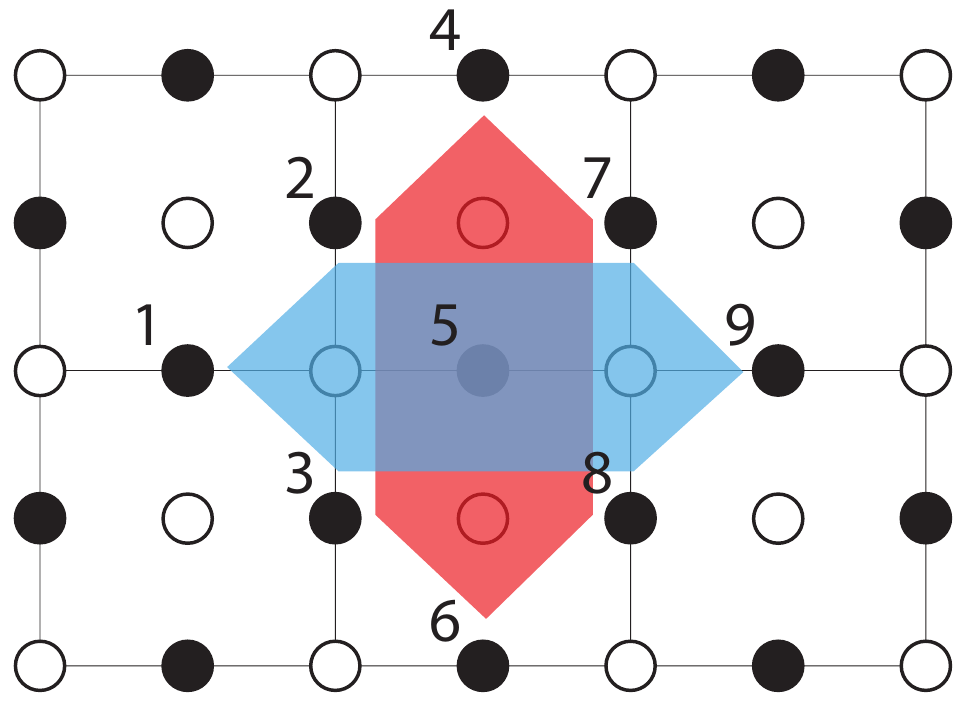}
  \caption{Surface code fragment to inject an arbitrary qubit.
  The lattice has only normal stabilizers at first.
  The shown superstabilizers are introduced in several steps, as described in Section~\ref{sec:state_injection}.
  }
  \label{fig:state_injection}
 \end{center}
\end{figure}

Direct conversion from a two-defect surface code qubit to a deformation-based qubit can be achieved.
This conversion works as the state injection for the deformation-based qubit and
e.g. to support
networking among multiple quantum computers that employ heterogeneous error correcting codes
~\cite{PhysRevA.93.042338}.
To complete universality of the deformation-based surface code, we demonstrate the arbitrary state injection in this section.
We first inject an arbitrary qubit to a two-defect surface code following Fowler et al.
~\cite{Fowler:2009High-threshold_universal_quantum_computation_on_the_surface_code},
as depicted on a fragment of surface code in Figure \ref{fig:state_injection}.
The surface begins in normal operation, using qubit 5 and
measuring all 4-qubit stabilizers,
\begin{equation}
\begin{array}{c|ccccccccc}
 &1&2 &3 &4 &5 &6 &7 &8 &9 \\
 \hline
 &X&X &X &  &X & & & & \\
 & &  &  &  &X &  &X &X &X \\
 & &Z &  &Z &Z &  &Z & & \\
 & &  &Z &  &Z &Z &  &Z & \\
\end{array}
\end{equation}
where each number corresponds to the number in Figure~\ref{fig:state_injection}.
First, we measure qubit 5 in the X basis,
disentangling it from the larger state where $M_a^b$ denotes a measured value where $a$ is the measurement basis and $b$ is the qubit index.
\begin{equation}
\begin{array}{c|ccccccccc}
 & 1&2 &3 &4 &5 &6 &7 &8 &9 \\
 \hline
 & X&X &X &  &X & & & & \\
 &  &  &  &  &X &  &X &X &X \\
(-1)^{M_X^5} &  &  &  &  &X &  &  & & \\
 &  &Z &Z &Z &  &Z &Z &Z & \\
\end{array}
\end{equation}
If the -1 eigenvalue is measured, apply either $Z_2Z_4Z_5Z_7$ or $Z_3Z_5Z_6Z_8$
to restore $X_1X_2X_3$ and $X_7X_8X_9$ to +1 eigenvalues,
\begin{equation}
\begin{array}{c|ccccccccc}
 & 1&2 &3 &4 &5 &6 &7 &8 &9 \\
 \hline
 & X&X &X &  & & & & & \\
 &  &  &  &  & &  &X &X &X \\
 &  &  &  &  &X &  &  & & \\
 &  &Z &Z &Z &  &Z &Z &Z &
\end{array}.
\end{equation}
Next, qubit 5 is rotated to the arbitrary desired state~\footnote{\emph{Note:} Eqs. 5-9 and 14-21 describe \emph{states} that are stabilized by the corresponding terms, but do not correspond directly to stabilizer measurements conducted for error correction purposes.  In particular, the last line of Eq. 8 and 9 represents the newly introduced superstabilizer itself, while the two stabilizers just above illustrate the degree of freedom representing our logical qubit.  The stabilizers demarking the degree of freedom are labeled in the leftmost column inside the parentheses with $+$ or $-$ as appropriate.}, $\alpha (Z) + \beta (-Z)$,
 \begin{eqnarray}
  \alpha \left(\begin{array}{c|ccccccccc}
 & 1&2 &3 &4 &5 &6 &7 &8 &9 \\
 \hline
 & X&X &X &  & & & & & \\
 &  &  &  &  & &  &X &X &X \\
 +&  &  &  &  &Z &  &  & & \\
 &  &Z &Z &Z &  &Z &Z &Z & \\
	  \end{array}  \right)& \nonumber\\
 + \beta \left(\begin{array}{c|ccccccccc}
 & 1&2 &3 &4 &5 &6 &7 &8 &9 \\
 \hline
 & X&X &X &  & & & & & \\
 &  &  &  &  & &  &X &X &X \\
 -&  &  &  &  &Z &  &  & & \\
 &  &Z &Z &Z &  &Z &Z &Z & \\
	  \end{array}  \right).
 \end{eqnarray}
 Then we measure $Z_2Z_4Z_5Z_7$ and $Z_3Z_5Z_6Z_8$,
  \begin{eqnarray}
  \alpha \left(\begin{array}{c|ccccccccc}
 & 1&2 &3 &4 &5 &6 &7 &8 &9 \\
 \hline
 & X&X &X &  & & &X &X &X \\
 +&  &  &  &  &Z &  &  & & \\
 (-1)^{M_Z^{2457}} &  &Z & &Z &Z & &Z & & \\
 (-1)^{M_Z^{3568}} &  & &Z & &Z &Z & &Z & \\
	  \end{array}  \right)& \nonumber\\
 + \beta \left(\begin{array}{c|ccccccccc}
 & 1&2 &3 &4 &5 &6 &7 &8 &9 \\
 \hline
 & X&X &X &  & & &X &X &X \\
 -&  &  &  &  &Z &  &  & & \\
 (-1)^{M_Z^{2457}} &  &Z & &Z &Z & &Z & & \\
 (-1)^{M_Z^{3568}} &  & &Z & &Z &Z & &Z & \\
	  \end{array}  \right).
 \end{eqnarray}
 If the -1 eigenvalue is measured, apply either $X_1X_2X_3$ or $X_7X_8X_9$
 to give the desired state.
 The two defects exist at $X_1X_2X_3X_5$ and $X_5X_7X_8X_9$,
 a minimal logical qubit of distance 1,
\begin{eqnarray}
  \alpha \left(\begin{array}{c|ccccccccc}
 & 1&2 &3 &4 &5 &6 &7 &8 &9 \\
 \hline
 & X&X &X &  & & &X &X &X \\
 +&  &  &  &  &Z &  &  & & \\
 &  &Z & &Z &Z & &Z & & \\
 &  &  &Z & &Z &Z & &Z & \\
	  \end{array}  \right)& \nonumber\\
 + \beta \left(\begin{array}{c|ccccccccc}
 & 1&2 &3 &4 &5 &6 &7 &8 &9 \\
 \hline
 & X&X &X &  & & &X &X &X \\
 -&  &  &  &  &Z &  &  & & \\
 &  &Z &  &Z &Z & &Z & & \\
 &  &  &Z &  &Z &Z & &Z & \\
	  \end{array}  \right).
\end{eqnarray}

So far we have the logical qubit of the two-defect surface code.
Next we start to convert this logical qubit 
to the deformation-based surface code.

For pedagogical clarity,
we omit writing the stabilizers that do not change over the course of this operation,
depicted in white in the figures, and
we write $Z_2Z_4Z_5Z_7 \otimes Z_3Z_5Z_6Z_8 = Z_2Z_3Z_4Z_6Z_7Z_8$, which is a product of two stabilizers
and which can be measured as a stabilizer without breaking the logical state.
We again measure qubit 5 in the $X$ basis,
merging the two minimal defects into one superstabilizer,
    \begin{eqnarray}
 \alpha \left(\begin{array}{c|ccccccccc}
 & 1&2 &3 &4 &5 &6 &7 &8 &9 \\
 \hline
 & X&X &X &  & & &X &X &X \\
(-1)^{M_X^5} &  &  &  &  &X &  &  & & \\
 +&  &Z &  &Z & & &Z & & \\
 +&  &  &Z &  & &Z & &Z & \\
 &  &Z &Z &Z & &Z &Z &Z & \\
	  \end{array}  \right)& \nonumber\\
 + \beta \left(\begin{array}{c|ccccccccc}
 & 1&2 &3 &4 &5 &6 &7 &8 &9 \\
 \hline
 & X&X &X &  & & &X &X &X \\
(-1)^{M_X^5} &  &  &  &  &X &  &  & & \\
- &  &Z &  &Z & & &Z & & \\
- &  &  &Z &  & &Z & &Z & \\
 &  &Z &Z &Z & &Z &Z &Z & \\
	  \end{array}  \right).
    \end{eqnarray}
    If the -1 eigenvalue is obtained,
    apply either $Z_2Z_4Z_5Z_7$ or $Z_3Z_5Z_6Z_8$
    to keep the logical $X$ operator such as $X_1X_2X_3X_5$ into $X_1X_2X_3$, giving
    
        \begin{eqnarray}
  \alpha \left(\begin{array}{c|ccccccccc}
 & 1&2 &3 &4 &5 &6 &7 &8 &9 \\
 \hline
 & X&X &X &  & & &X &X &X \\
 &  &  &  &  &X &  &  & & \\
 +&  &Z &  &Z & & &Z & & \\
 +&  &  &Z &  & &Z & &Z & \\
 &  &Z &Z &Z & &Z &Z &Z & \\
	  \end{array}  \right)& \nonumber\\
 + \beta \left(\begin{array}{c|ccccccccc}
 & 1&2 &3 &4 &5 &6 &7 &8 &9 \\
 \hline
 & X&X &X &  & & &X &X &X \\
 &  &  &  &  &X &  &  & & \\
- &  &Z &  &Z & & &Z & & \\
- &  &  &Z &  & &Z & &Z & \\
 &  &Z &Z &Z & &Z &Z &Z & \\
	  \end{array}  \right).
    \end{eqnarray}

    Now $Z_2Z_4Z_7$ and $Z_3Z_6Z_8$ share the desired state.
    We can now begin measuring $Z_2Z_3Z_4Z_6Z_7Z_8$ as our superstabilizer.
    As is common with state injections,
    because the process begins with a raw qubit,
    state distillation on the logical qubit is required after this process.

\section{CNOT gate}
\label{sec:cnot}
A CNOT gate can be performed utilizing lattice surgery~\cite{Horsman:2012lattice_surgery}.
The basic concept of the CNOT gate by lattice surgery is
\begin{enumerate}
 \item prepare a control (C) qubit in $\alpha \vert 0_{C} \rangle + \beta \vert 1_{C} \rangle$ and
	 a target (T) qubit in $\alpha' \vert 0_{T} \rangle + \beta' \vert 1_{T} \rangle$.
 \item prepare an intermediate (INT) qubit in $\vert +_{I} \rangle$.
	 The initial state is
\begin{equation}
	 \vert \psi ^{init} \rangle = (\alpha \vert 0_{C} \rangle + \beta \vert 1_{C} \rangle) \otimes \vert +_{I} \rangle \otimes (\alpha' \vert 0_{T} \rangle + \beta' \vert 1_{T} \rangle).
\end{equation}
 \item measure $Z_CZ_{I}$ and get
\begin{equation}
	 \vert \psi' \rangle = (\alpha \vert 0_{C}0_{I} \rangle + \beta \vert 1_{C}1_{I} \rangle) \otimes (\alpha' \vert 0_{T} \rangle + \beta' \vert 1_{T} \rangle)
\end{equation}
	 by applying $X_{I}$ if the -1 eigenvalue is observed.
 \item measure $X_{I}X_T$ and get
 \begin{eqnarray}
  \vert \psi '' \rangle = 
	 \alpha \vert 0_{C} \rangle (
	 \alpha' \vert 0_{I}0_{T} \rangle + \beta' \vert 0_{I}1_{T} \rangle + \beta' \vert 1_{I}0_{T} \rangle + \alpha' \vert 1_{I}1_{T} \rangle) \nonumber\\
	 + \beta \vert 1_{C} \rangle (
	  \beta' \vert 0_{I}0_{T} \rangle + \alpha' \vert 0_{I}1_{T} \rangle + \alpha' \vert 1_{I}0_{T} \rangle + \beta' \vert 1_{I}1_{T} \rangle	 ) \nonumber\\
	  \label{equ:ls}
 \end{eqnarray}
	 if the +1 eigenvalue is observed, and get
	 \begin{eqnarray}
 \vert \psi ''' \rangle = 
	 \alpha \vert 0_{C} \rangle (
	 \alpha' \vert 0_{I}0_{T} \rangle + \beta' \vert 0_{I}1_{T} \rangle - \beta' \vert 1_{I}0_{T} \rangle - \alpha' \vert 1_{I}1_{T} \rangle	 ) \nonumber\\
	 + \beta \vert 1_{C} \rangle (
	 - \beta' \vert 0_{I}0_{T} \rangle - \alpha' \vert 0_{I}1_{T} \rangle + \alpha' \vert 1_{I}0_{T} \rangle + \beta' \vert 1_{I}1_{T} \rangle) \nonumber\\
	 \end{eqnarray}
	 if the -1 eigenvalue is observed. Apply $Z_CZ_I$ and get Equation \ref{equ:ls} when -1 is observed.
	 Merging $I$ and $T$ by the lattice surgery, the Z operators are XORed and finally we get
	 \begin{eqnarray}
 \vert \psi ^{final} \rangle = 
	 	 \alpha \vert 0_{C} \rangle (
		  \alpha' \vert 0_{m} \rangle + \beta' \vert 1_{m} \rangle) 
		  + \beta \vert 1_{C} \rangle (\beta' \vert 0_{m} \rangle + \alpha' \vert 1_{m} \rangle) \nonumber\\
	 \end{eqnarray}
	 where \textit{m} stands for \textit{merged}, indicating the merged qubit of $I$ and $T$.
\end{enumerate}

\begin{figure}[t]
 \begin{center}
  \includegraphics[width=8cm]{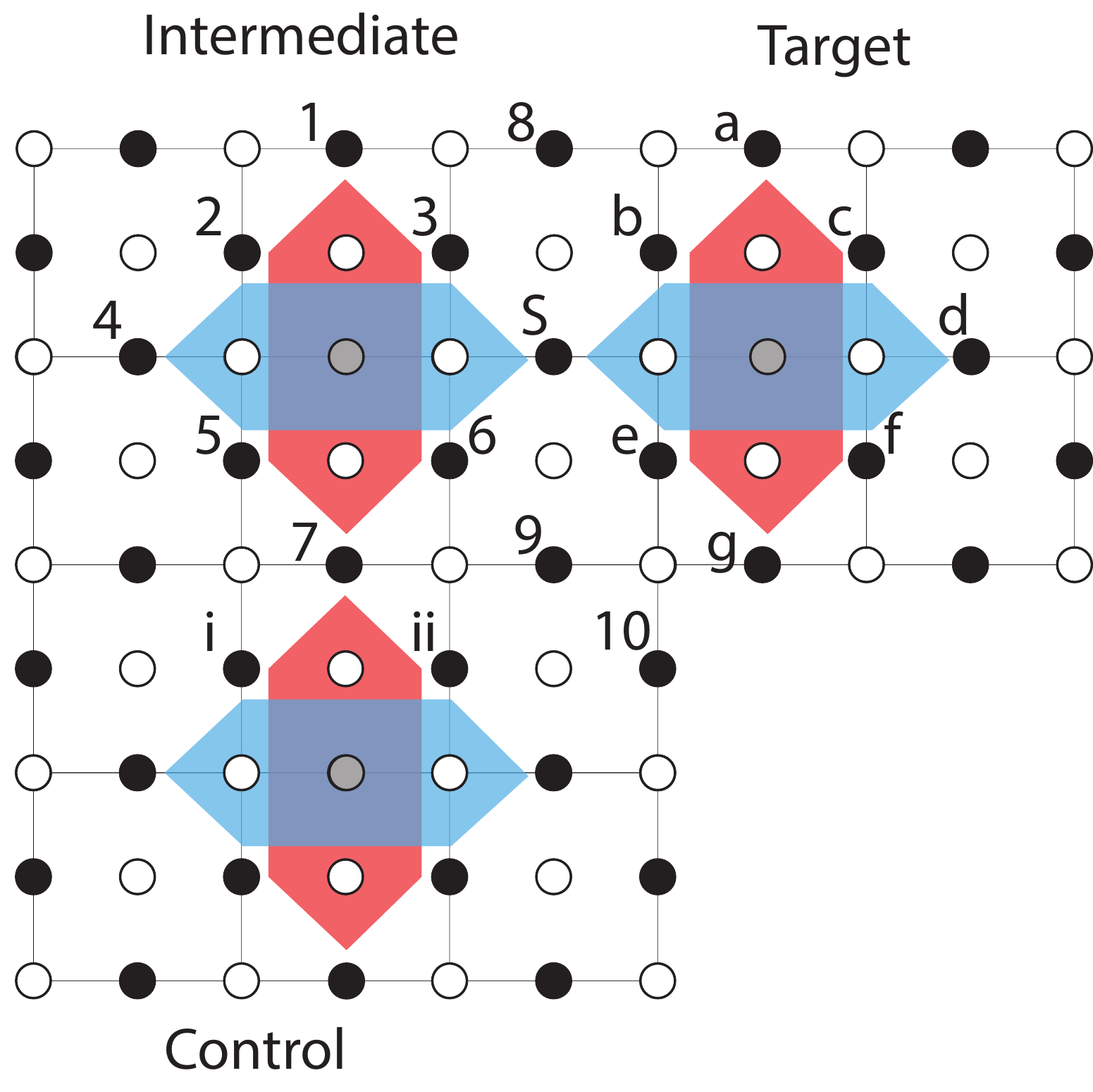}
  \caption{
  Three deformation-based qubits to demonstrate CNOT gate between the control qubit and the target qubit by lattice-surgery like operations
  in Section \ref{sec:cnot}.
  The intermediate qubit is initialized in $\vert + \rangle$.
  The code distance for those logical qubits is still 3 during lattice surgery.
  }
  \label{fig:lattice_surgery}
 \end{center}
\end{figure}

Figure~\ref{fig:lattice_surgery} depicts the logical CNOT gate of the deformation-based qubit by lattice surgery.

To measure $Z_CZ_{I}$, we measure $Z_5Z_6Z_{i}Z_{ii}$. This is achieved by swapping qubit $7$ with a neighboring ancilla qubit and
using the fault-tolerant stabilizer measurement described in Section~\ref{sec:fast_stabilizer}.
This measurement is repeated $d$ times for majority voting to correct errors, where $d$ is the code distance.
If the -1 eigenvalue is observed from the $Z_CZ_{I}$ measurement, $X_{I}$ is applied.
During the measurement of $Z_CZ_{I}$, we cannot measure the $Z$ superstabilizers of the intermediate qubit and the control qubit, meanwhile normal $Z$ stabilizers can be measured.
Hence error chains connecting the two $Z$ superstabilizers, such as $X_7$ and $X_6X_9X_{ii}$
may be caused. (Figure ~\ref{sec:cnot} shows distance 3 code, therefore we should not assume an error chain of length 3.)
However, those error chains do not matter since they are stabilizers for $Z_5Z_6Z_{i}Z_{ii}$.

Next, we measure $X_{I}X_T$ and merge the intermediate qubit and the target qubit.
Here we describe the merge operation of deformation-based qubits.
The original state is
\begin{eqnarray}
&&(\alpha \vert 0_{C}0_{I} \rangle + \beta \vert 1_{C}1_{I} \rangle) \otimes (\alpha' \vert 0_{T} \rangle + \beta' \vert 1_{T} \rangle) \nonumber \\
 &=&\alpha\alpha'\vert 0_{C}0_{I}0_{T} \rangle +\alpha\beta'\vert 0_{C}0_{I}1_{T} \rangle \nonumber \\
 &&+\beta\alpha'\vert 1_{C}1_{I}0_{T} \rangle +\beta\beta'\vert 1_{C}1_{I}1_{T} \rangle
\label{equ:merge_orig_state}
 .
\end{eqnarray}
The first term of Equation~\ref{equ:merge_orig_state} is

         \begin{equation}
  \alpha\alpha' \vert0_C\rangle \left(\begin{array}{p{1.5mm}|p{1.5mm}p{1.5mm}p{1.5mm}p{1.5mm}p{1.5mm}p{1.5mm}p{1.5mm}p{1.5mm}p{1.5mm}p{1.5mm}p{1.5mm}p{1.5mm}p{1.5mm}p{1.5mm}p{1.5mm}p{1.5mm}p{1.5mm}}
	    &1&2&3&4&5&6&7&8&9&S&a&b&c&d&e&f&g\\
		    \hline
&$Z$&$Z$&$Z$&&$Z$&$Z$&$Z$&&&&&&&&&&\\
&&$X$&$X$&$X$&$X$&$X$&&&&$X$&&&&&&&\\
&&&&&&&&&&&$Z$&$Z$&$Z$&&$Z$&$Z$&$Z$\\
&&&&&&&&&&$X$&&$X$&$X$&$X$&$X$&$X$&\\
&&&$Z$&&&&&$Z$&&$Z$&&$Z$&&&&&\\
&&&&&&$Z$&&&$Z$&$Z$&&&&&$Z$&&\\
$+$&&&&&$Z$&$Z$&$Z$&&&&&&&&&&\\
$+$&&&&&&&&&&&&&&&$Z$&$Z$&$Z$\\
	  \end{array}  \right)
	   \end{equation}

where the logical state of two qubits exists in $Z_1Z_2Z_3$ and $Z_aZ_bZ_c$.
The two bottom lines are the logical operator states.
Measure qubit $S$ in the $Z$ basis, giving
         \begin{equation}
  \alpha\alpha' \vert0_C\rangle \left(\begin{array}{c|p{1.5mm}p{1.5mm}p{1.5mm}p{1.5mm}p{1.5mm}p{1.5mm}p{1.5mm}p{1.5mm}p{1.5mm}p{1.5mm}p{1.5mm}p{1.5mm}p{1.5mm}p{1.5mm}p{1.5mm}p{1.5mm}p{1.5mm}}
	    &1&2&3&4&5&6&7&8&9&S&a&b&c&d&e&f&g\\
		    \hline
&$Z$&$Z$&$Z$&&$Z$&$Z$&$Z$&&&&&&&&&&\\
&&$X$&$X$&$X$&$X$&$X$&&&&&&$X$&$X$&$X$&$X$&$X$&\\
&&&&&&&&&&&$Z$&$Z$&$Z$&&$Z$&$Z$&$Z$\\
(-1)^{M_Z^S}&&&$Z$&&&&&$Z$&&&&$Z$&&&&&\\
(-1)^{M_Z^S}&&&&&&$Z$&&&$Z$&&&&&&$Z$&&\\
(-1)^{M_Z^S}&&&&&&&&&&$Z$&&&&&&&\\
+&&&&&$Z$&$Z$&$Z$&&&&&&&&&&\\
+&&&&&&&&&&&&&&&$Z$&$Z$&$Z$\\
	  \end{array}  \right).
	   \end{equation}

If -1 eigenvalue is obtained, apply either $X_2X_3X_4X_5X_6X_S$ or $X_bX_cX_dX_eX_fX_S$ and get

         \begin{equation}
  \alpha\alpha' \vert0_C\rangle \left(\begin{array}{p{1.5mm}|p{1.5mm}p{1.5mm}p{1.5mm}p{1.5mm}p{1.5mm}p{1.5mm}p{1.5mm}p{1.5mm}p{1.5mm}p{1.5mm}p{1.5mm}p{1.5mm}p{1.5mm}p{1.5mm}p{1.5mm}p{1.5mm}p{1.5mm}}
	    &1&2&3&4&5&6&7&8&9&S&a&b&c&d&e&f&g\\
		    \hline
&$Z$&$Z$&$Z$&&$Z$&$Z$&$Z$&&&&&&&&&&\\
&&$X$&$X$&$X$&$X$&$X$&&&&&&$X$&$X$&$X$&$X$&$X$&\\
&&&&&&&&&&&$Z$&$Z$&$Z$&&$Z$&$Z$&$Z$\\
&&&$Z$&&&&&$Z$&&&&$Z$&&&&&\\
&&&&&&$Z$&&&$Z$&&&&&&$Z$&&\\
&&&&&&&&&&$Z$&&&&&&&\\
$+$&&&&&$Z$&$Z$&$Z$&&&&&&&&&&\\
$+$&&&&&&&&&&&&&&&$Z$&$Z$&$Z$\\
	  \end{array}  \right).
	   \end{equation}

	   Next, we measure $X_3X_bX_6X_e$ for the third step of lattice surgery.
	   We can measure $X_3$, $X_b$, $X_6$ and $X_e$ both to execute our merge and to measure $X_3X_bX_6X_e$.
	   Measure qubit $3$ in the $X$ basis. If -1 is obtained, apply either $Z_3Z_8Z_b$ or $Z_1Z_2Z_3Z_5Z_6Z_7$.

         \begin{equation}
  \alpha\alpha' \vert0_C\rangle \left(\begin{array}{p{1.5mm}|p{1.5mm}p{1.5mm}p{1.5mm}p{1.5mm}p{1.5mm}p{1.5mm}p{1.5mm}p{1.5mm}p{1.5mm}p{1.5mm}p{1.5mm}p{1.5mm}p{1.5mm}p{1.5mm}p{1.5mm}p{1.5mm}p{1.5mm}}
	    &1&2&3&4&5&6&7&8&9&S&a&b&c&d&e&f&g\\
		    \hline
&$Z$&$Z$&&&$Z$&$Z$&$Z$&$Z$&&&&$Z$&&&&&\\
&&$X$&&$X$&$X$&$X$&&&&&&$X$&$X$&$X$&$X$&$X$&\\
&&&&&&&&&&&$Z$&$Z$&$Z$&&$Z$&$Z$&$Z$\\
&&&&&&$Z$&&&$Z$&&&&&&$Z$&&\\
$+$&&&&&$Z$&$Z$&$Z$&&&&&&&&&&\\
$+$&&&&&&&&&&&&&&&$Z$&$Z$&$Z$\\
	  \end{array}  \right)
	   \end{equation}

Measure qubit $b$ in the $X$ basis. If the -1 is obtained, apply either $Z_1Z_2Z_5Z_6Z_7Z_8Z_b$ or $Z_aZ_bZ_cZ_eZ_fZ_g$.

         \begin{equation}
  \alpha\alpha' \vert0_C\rangle \left(\begin{array}{p{1.5mm}|p{1.5mm}p{1.5mm}p{1.5mm}p{1.5mm}p{1.5mm}p{1.5mm}p{1.5mm}p{1.5mm}p{1.5mm}p{1.5mm}p{1.5mm}p{1.5mm}p{1.5mm}p{1.5mm}p{1.5mm}p{1.5mm}p{1.5mm}}
	    &1&2&3&4&5&6&7&8&9&S&a&b&c&d&e&f&g\\
		    \hline
&$Z$&$Z$&&&$Z$&$Z$&$Z$&$Z$&&&$Z$&&$Z$&&$Z$&$Z$&$Z$\\
&&$X$&&$X$&$X$&$X$&&&&&&&$X$&$X$&$X$&$X$&\\
&&&&&&$Z$&&&$Z$&&&&&&$Z$&&\\
$+$&&&&&$Z$&$Z$&$Z$&&&&&&&&&&\\
$+$&&&&&&&&&&&&&&&$Z$&$Z$&$Z$\\
	  \end{array}  \right)
	   \end{equation}

Measure qubit $6$ in the $X$ basis, and apply either $Z_1Z_2Z_5Z_6Z_7Z_8Z_aZ_cZ_eZ_fZ_g$ if the -1 eigenvalue is observed.

         \begin{equation}
  \alpha\alpha' \vert0_C\rangle \left(\begin{array}{p{1.5mm}|p{1.5mm}p{1.5mm}p{1.5mm}p{1.5mm}p{1.5mm}p{1.5mm}p{1.5mm}p{1.5mm}p{1.5mm}p{1.5mm}p{1.5mm}p{1.5mm}p{1.5mm}p{1.5mm}p{1.5mm}p{1.5mm}p{1.5mm}}
	    &1&2&3&4&5&6&7&8&9&S&a&b&c&d&e&f&g\\
		    \hline
&$Z$&$Z$&&&$Z$&&$Z$&$Z$&$Z$&&$Z$&&$Z$&&&$Z$&$Z$\\
&&$X$&&$X$&$X$&&&&&&&&$X$&$X$&$X$&$X$&\\
$+$&&&&&$Z$&&$Z$&&$Z$&&&&&&$Z$&&\\
$+$&&&&&&&&&&&&&&&$Z$&$Z$&$Z$\\
	  \end{array}  \right)
	   \end{equation}

Measure qubit $e$ in the $X$ basis and apply both $Z_5Z_7Z_9$ as $Z_I$ and $Z_iZ_7Z_{ii}$ as $Z_C$ if the -1 eigenvalue is obtained.

Alternately, we can measure $X_3$, $X_b$, $X_6$ and $X_e$ in parallel.
After the parallel measurements, if an even number of $-1$ eigenvalues is observed, as in normal error correction,
a physical $Z$ operator chain connecting the remaining $X$ stabilizers with $-1$ eigenvalues is executed.
If an odd number of $-1$ eigenvalue is observed,
we execute the physical $Z$ operator chain and there still remains an $X$ stabilizer with $-1$ eigenvalue.
The $X$ superstabilizer of the merged qubit actually has the $-1$ eigenvalue in this case, hence we connect
the remaining $X$ stabilizer and the intermediate qubit side of the $X$ superstabilizer.
This operation keeps the eigenvalues of the lattice $+1$ and works as $Z_I$, like $Z_5Z_7Z_9$ was used in the sequential form above.
We execute $Z_iZ_7Z_{ii}$ as $Z_C$ when an odd number of $-1$ eigenvalue is observed.

Those measurements work for connecting the superstabilizers.
Therefore, those measurements are allowed to be non-fault-tolerant since the remaining stabilizers confirm the correctness of the measurements;
when qubit $e$ is measured in the $X$ basis, regardless of whether a measurement error occurs, if the remaining stabilizer $X_9X_gX_{10}$ outputs -1 repeatedly, we can conclude the correct measurement of qubit $e$ to be -1.

Now we have code space for only one qubit and the two qubits are merged into a qubit
whose logical operator state is the product of the first two, shown in the bottom line,
         \begin{equation}
  \alpha\alpha' \vert0_C\rangle \left(\begin{array}{p{1.5mm}|p{1.5mm}p{1.5mm}p{1.5mm}p{1.5mm}p{1.5mm}p{1.5mm}p{1.5mm}p{1.5mm}p{1.5mm}p{1.5mm}p{1.5mm}p{1.5mm}p{1.5mm}p{1.5mm}p{1.5mm}p{1.5mm}p{1.5mm}}
	    &1&2&3&4&5&6&7&8&9&S&a&b&c&d&e&f&g\\
		    \hline
&$Z$&$Z$&&&$Z$&&$Z$&$Z$&$Z$&&$Z$&&$Z$&&&$Z$&$Z$\\
&&$X$&&$X$&$X$&&&&&&&&$X$&$X$&&$X$&\\
$+$&&&&&$Z$&&$Z$&&$Z$&&&&&&&$Z$&$Z$\\
	  \end{array}  \right).
	   \end{equation}

By similar operations, Equation~\ref{equ:merge_orig_state} is rewritten to
 \begin{flushleft}
  \begin{eqnarray}
  \alpha\alpha' \vert0_C\rangle \left(\begin{array}{p{1.5mm}|p{1.5mm}p{1.5mm}p{1.5mm}p{1.5mm}p{1.5mm}p{1.5mm}p{1.5mm}p{1.5mm}p{1.5mm}p{1.5mm}p{1.5mm}p{1.5mm}p{1.5mm}p{1.5mm}p{1.5mm}p{1.5mm}p{1.5mm}}
	    &1&2&3&4&5&6&7&8&9&S&a&b&c&d&e&f&g\\
		    \hline
&$Z$&$Z$&&&$Z$&&$Z$&$Z$&$Z$&&$Z$&&$Z$&&&$Z$&$Z$\\
&&$X$&&$X$&$X$&&&&&&&&$X$&$X$&&$X$&\\
$+$&&&&&$Z$&&$Z$&&$Z$&&&&&&&$Z$&$Z$\\
	  \end{array}  \right)\nonumber\\
+  \alpha\beta' \vert0_C\rangle \left(\begin{array}{p{1.5mm}|p{1.5mm}p{1.5mm}p{1.5mm}p{1.5mm}p{1.5mm}p{1.5mm}p{1.5mm}p{1.5mm}p{1.5mm}p{1.5mm}p{1.5mm}p{1.5mm}p{1.5mm}p{1.5mm}p{1.5mm}p{1.5mm}p{1.5mm}}
	    &1&2&3&4&5&6&7&8&9&S&a&b&c&d&e&f&g\\
		    \hline
&$Z$&$Z$&&&$Z$&&$Z$&$Z$&$Z$&&$Z$&&$Z$&&&$Z$&$Z$\\
&&$X$&&$X$&$X$&&&&&&&&$X$&$X$&&$X$&\\
$-$&&&&&$Z$&&$Z$&&$Z$&&&&&&&$Z$&$Z$\\
	  \end{array}  \right)\nonumber\\
+  \beta\alpha' \vert1_C\rangle \left(\begin{array}{p{1.5mm}|p{1.5mm}p{1.5mm}p{1.5mm}p{1.5mm}p{1.5mm}p{1.5mm}p{1.5mm}p{1.5mm}p{1.5mm}p{1.5mm}p{1.5mm}p{1.5mm}p{1.5mm}p{1.5mm}p{1.5mm}p{1.5mm}p{1.5mm}}
	    &1&2&3&4&5&6&7&8&9&S&a&b&c&d&e&f&g\\
		    \hline
&$Z$&$Z$&&&$Z$&&$Z$&$Z$&$Z$&&$Z$&&$Z$&&&$Z$&$Z$\\
&&$X$&&$X$&$X$&&&&&&&&$X$&$X$&&$X$&\\
$-$&&&&&$Z$&&$Z$&&$Z$&&&&&&&$Z$&$Z$\\
	  \end{array}  \right)\nonumber\\
+  \beta\beta' \vert1_C\rangle \left(\begin{array}{p{1.5mm}|p{1.5mm}p{1.5mm}p{1.5mm}p{1.5mm}p{1.5mm}p{1.5mm}p{1.5mm}p{1.5mm}p{1.5mm}p{1.5mm}p{1.5mm}p{1.5mm}p{1.5mm}p{1.5mm}p{1.5mm}p{1.5mm}p{1.5mm}}
	    &1&2&3&4&5&6&7&8&9&S&a&b&c&d&e&f&g\\
		    \hline
&$Z$&$Z$&&&$Z$&&$Z$&$Z$&$Z$&&$Z$&&$Z$&&&$Z$&$Z$\\
&&$X$&&$X$&$X$&&&&&&&&$X$&$X$&&$X$&\\
$+$&&&&&$Z$&&$Z$&&$Z$&&&&&&&$Z$&$Z$\\
	  \end{array}  \right).\nonumber\\
\label{eqn:merge_2}
 \end{eqnarray}
 \end{flushleft}
 Using a new definition, we now have
 \begin{eqnarray}
  \vert 0_{m} \rangle = Z_5Z_7Z_9Z_fZ_g\\
  \vert 1_{m} \rangle = -Z_5Z_7Z_9Z_fZ_g
 \end{eqnarray}
 where $m$ stands for $merged$. Equation~\ref{eqn:merge_2} can be written as
 \begin{equation}
  \alpha\vert0\rangle(\alpha'\vert0_{m}\rangle + \beta'\vert1_{m}\rangle) + \beta\vert1\rangle(\beta'\vert0_{m}\rangle + \alpha'\vert1_{m}\rangle)
 \end{equation}
therefore now we have a complete CNOT gate.
From this point in the operation,
we start to measure the new superstabilizers.

\section{Arbitrary size stabilizer measurement}
\label{sec:fast_stabilizer}
We suggest using a cat state of an arbitrary length to measure superstabilizers.
%
%
In this section, we first discuss fault-tolerant preparation, then generic use of cat states for constant-time stabilizer measurement, before addressing superstabilizers in our system.
Finally, we return to the issue of errors.

\subsection{Arbitrary length cat state preparation}
\begin{figure}[t]
  \begin{center}
   \includegraphics[width=6cm]{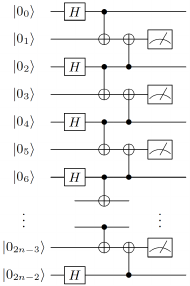}
  \end{center}
  \caption{Non-fault-tolerant circuit to make a $n$-size cat state in 5 steps.}
  \label{fig:cat_state_creation}
 \end{figure}

The non-fault-tolerant circuit to prepare an arbitrary length cat state in constant time is depicted in Figure~\ref{fig:cat_state_creation}.
In the circuit, many qubits in $\vert + \rangle$ are created and entangled by measuring $ZZ$ of every pair of neighboring qubits.
Here, we prepare two qubits in $\vert +_0+_2\rangle$ and a third qubit in $\vert 0_1 \rangle$,
\begin{equation}
 \vert \psi_{012} \rangle = \vert +_0 0_1 +_2 \rangle,
\end{equation}
with this order corresponding to the physical placement.
Dispensing with normalization, as every term has the same amplitude,
apply CNOT for $Z_0Z_2$ measurement:
\begin{align}
 \vert \psi_{012}' \rangle &= CNOT[2,1]CNOT[0,1] \vert +_0 0_1 +_2 \rangle \nonumber \\
 &= \vert 0_00_10_2 \rangle + \vert 0_01_11_2 \rangle + \vert 1_01_10_2 \rangle + \vert 1_00_11_2 \rangle 
\end{align}
where $CNOT[a,b]$ denotes that qubit $a$ is the control qubit and $b$ is the target.
Measure the ancilla qubit $1$ in the $Z$ basis and if the $-1$ eigenvalue is obtained, apply $X_1$ to get
\begin{equation}
\vert\psi_{02}''\rangle = \vert 0_00_2 \rangle + \vert 1_01_2 \rangle.
\end{equation}
We can entangle another qubit in $\vert + \rangle$ to this state in the same way and 
we can make a cat state of arbitrary length.
However, this procedure is not fault-tolerant and there is a chance of getting a problematic state
such as $\vert 00001111 \rangle + \vert 11110000 \rangle$. Using this state for a stabilizer measurement may produce a logical error
because the logical operator of the deformation-based qubit is a half of a superstabilizer.
Therefore we need to confirm that we have a proper cat state.
It is well-known that measuring $ZZ$ of every pair of qubits comprising the cat state
is good enough for this proof~\cite{nielsen-chuang:qci}.
Since measuring $ZZ$ of every pair of qubits requires many SWAP gates and a lot of steps,
we suggest repeating the $ZZ$ measurement of every pair of neighboring qubits $d-1$ times,
which guarantees the state is not in problematic state.
(The state yet could be an imperfect cat state such as $\vert 00100000 \rangle + \vert 11011111 \rangle$
due to individual physical errors, which is tolerable.)
 
 \subsection{Stabilizer measurement in constant time using cat state}
A three qubit cat state can be rewritten as
 \begin{align}
  \vert \psi_{cat}\rangle = &\vert 000 \rangle + \vert 111 \rangle \nonumber\\
  =& (\vert + \rangle + \vert - \rangle)(\vert + \rangle + \vert - \rangle)(\vert + \rangle + \vert - \rangle)\nonumber\\
  & + (\vert + \rangle - \vert - \rangle)(\vert + \rangle - \vert - \rangle)(\vert + \rangle - \vert - \rangle)\label{equ:cat_0}\\
  =& \vert +++ \rangle + \vert +-- \rangle + \vert -+- \rangle + \vert --+ \rangle.
  \label{equ:cat_1}
 \end{align}

 The $\vert 000 \rangle$ and $\vert 111 \rangle$ are rewritten in symmetric fashion
 except that the signs of factors involving an odd number of $\vert-\rangle$ differs, as shown in Equation~\ref{equ:cat_0}.
 From this fact and the binomial expansion, a cat state of any length involves an even number of $\vert-\rangle$.
 Applying a $Z$ to any qubit in the cat state, the state in Equation~\ref{equ:cat_1} is changed to 
\begin{equation}
\vert \psi_{cat}' \rangle = \vert -++ \rangle + \vert --- \rangle + \vert ++- \rangle + \vert +-+ \rangle.
\end{equation}
Applying a $Z$ to any qubit again, this state returns to the state in Equation~\ref{equ:cat_1}.
To observe whether we have the ``even'' cat state or the ``odd'' cat state,
we need to measure all qubits in the $X$ basis and calculate the product of the measured values.

Let us assume that
we have as many ancillae for the cat state as we have 
data qubits to stabilize, and we can assign a qubit in the cat state to each data qubit,
then apply CNOT from each cat state qubit to the corresponding data qubit.
This set of CNOTs is equivalent to the syndrome propagation for the $X_1X_2...X_n$ stabilizer.
The cat state starts from the ``even'' state and if an odd number of flip is performed the cat state results in the ``odd'' state.
The CNOTs can be applied simultaneously and the measurement can be performed simultaneously,
therefore this procedure requires three steps (CNOT, Hadamard and measurement in $Z$ basis).
  
\subsection{Superstabilizer implementation}
\begin{figure}[t]
   \begin{center}
    \includegraphics[width=8cm]{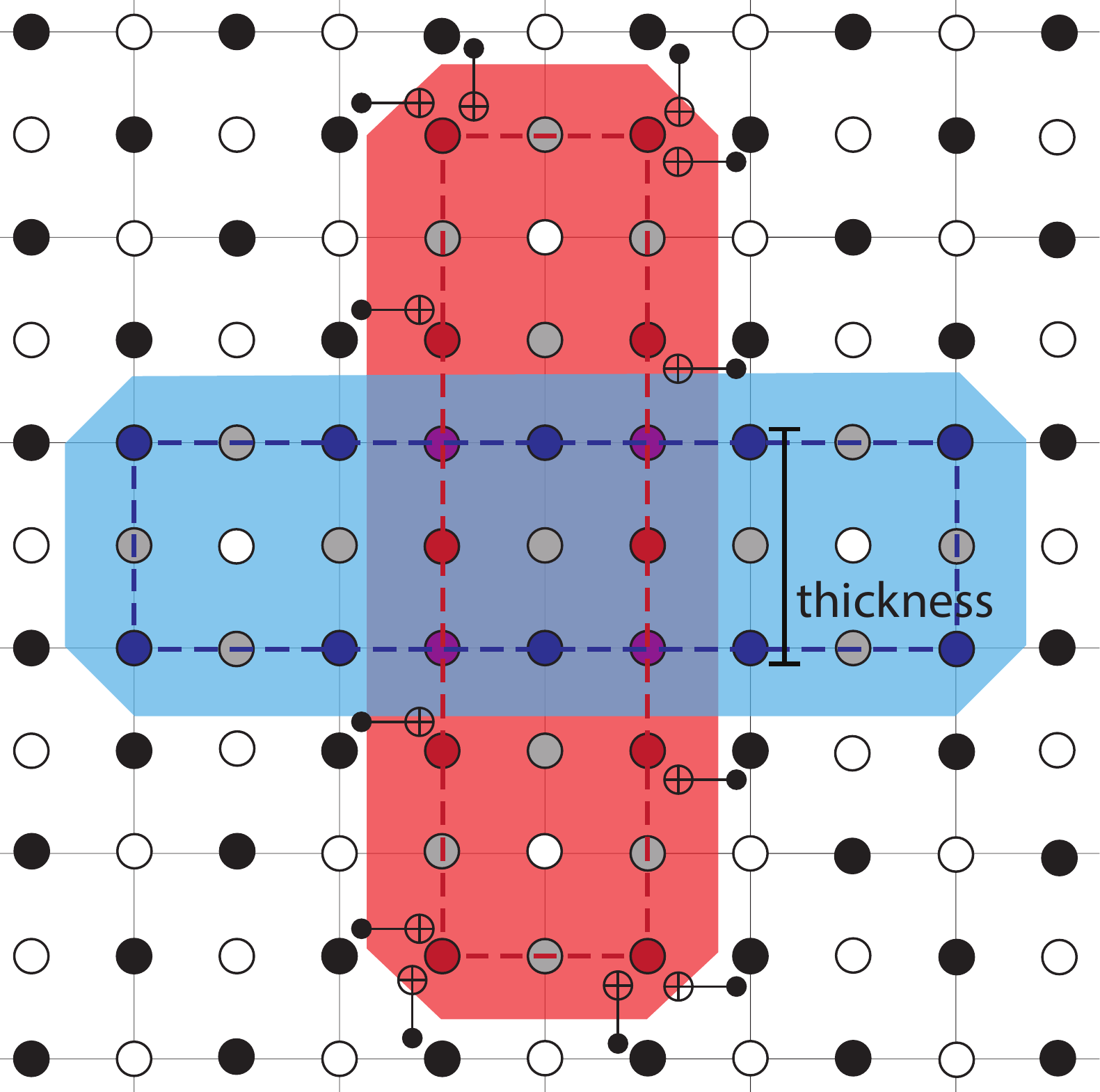}
    \caption{Implementation of two cat states for a superstabilizer.
    The red dots are ancilla qubits prepared in a cat state for the $Z$ superstabilizer.
    The red dashed loop describes the pairs for $ZZ$ stabilizers to create and confirm the cat state.
    The $ZZ$ stabilizer on each pair of neighboring red dots in this red dashed loop is executed.
    The gray qubits under the red dashed circle are qubits with odd indices in Figure~\ref{fig:cat_state_creation},
    used to measure $ZZ$ stabilizers.
    So as the blue dots and the blue dashed circle for the $X$ superstabilizer.
    The dots under the crosses of the dashed circles are used for both cat state creation alternately.
    The ``thickness'' of this deformation-based qubit is 2.
    The CNOT gates of the $Z$ superstabilizer are shown. Each ancilla qubit on the corner of the loop
    handles two data qubits and those along the sides handle one.
    }
    \label{fig:superstab}
   \end{center}
  \end{figure}

  The $ZZ$ stabilizers to confirm that we have a proper cat state, operated in a linear fashion, must be repeated $d$ times
  to suppress the probability of improper cat state generation to $O(p^{\lceil\frac{d}{2}\rceil})$,
  where $d$ is the code distance and $p$ is the physical error rate;
  A cat state placed in a circular fashion requires repeating $ZZ$ stabilizer $\lceil\frac{d}{2}\rceil$ times
  since $\lceil\frac{d}{2}\rceil$ physical errors are required to cause a logical error.
  Figure ~\ref{fig:superstab} depicts the placement of two sets of ancilla qubits,
  each of which is prepared in a cat state for the $X$ superstabilizer and for the $Z$ superstabilizer.
  The dashed lines describe the cat state qubits; red (blue) dots are qubits
  composing the cat state for the $Z$ ($X$) superstabilizer and
  gray dots are ancilla qubits to create and confirm the cat state (the ancillas' ancilla).
  The qubits under both dashed lines are used for the $Z$ and $X$ ancilla qubits alternately.
  Therefore we need $\frac{d}{2} \times 2 = d$ cycles to measure both the $Z$ superstabilizer and the $X$ superstabilizer.
  The ``thickness'' of the deformation-based qubit in Figure~\ref{fig:superstab} is 2 to allow us to have the loop cat state.
  Greater thickness requires fewer cycles of repeating $ZZ$ stabilizer to confirm the cat state.
  We assume that the thickness is 2 through the rest of this paper to show the basic idea of our architecture.

  The depth of the circuit to initialize a cat state is five.
  A cycle of the following $ZZ$ measurements for the proof requires four steps.
  The maximum number of \textsc{CNOT}s to propagate error syndromes from data qubits to an ancilla qubit is 2,
  as shown in Figure ~\ref{fig:superstab}, at the corners of the superstabilizers.
  The total number of steps to measure a superstabilizer is the sum of 
  $5 + 4 (d-1)=4d+1$ steps for cat state creation and the proof,
  $1$ step for a Hadamard gate for $Z$ superstabilizer,
  $2$ steps for syndrome propagation,
  $1$ step for a Hadamard gate for $X$ superstabilizer,
  $1$ step for measurements, where $d$ is the $code{\ }distance$.
  Therefore the number of steps to measure a superstabilizer is $4 d + 5$.

  \begin{figure}[t]
   \begin{center}
    \includegraphics[width=8cm]{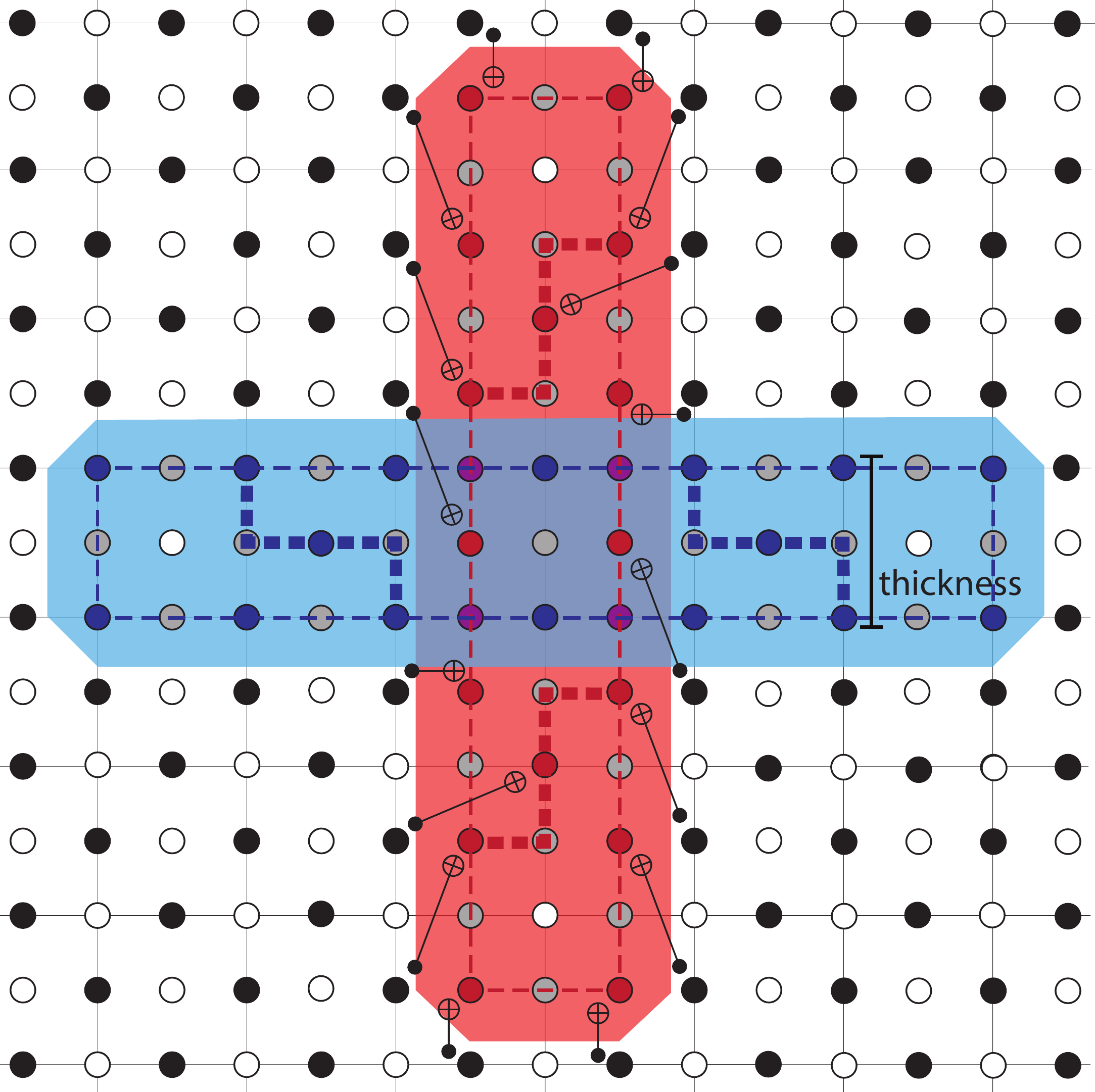}
    \caption{
    $Z$ superstabilizer in which a cat state qubit stabilizes a data qubit.
    Qubits on the thick dashed lines are newly added to the cat state qubits.
    It does not matter that cat state qubit on the cross of a thick dashed line and a thin dashed line
    is stabilized by three stabilizers for the proof of the correctness of the cat state
    since one cycle of stabilization for the proof takes four steps.
    Non-neighbor CNOT gates are executed after SWAP gates to neighbor the control and the target qubits.
    }
    \label{fig:superstab2}
   \end{center}
  \end{figure}
  By judicious use of the now-unused qubits in Figure \ref{fig:superstab}, we can recover the code distance lost in Figure \ref{fig:superstab2}.
For simplicity, we show thickness $t=2$ employing a cat state forming a complete loop,
in which each corner cat state qubit stabilizes two data qubits, resulting in reducing the effective code distance by $2$.
However, two data qubits neighboring a corner of a loop cat state have CNOT gates with the corner qubit
so that an error on the corner qubit may propagate to the two data qubits,
which may reduce the error suppression ability of the code.
Figure \ref{fig:superstab2} shows that, by utilizing unused physical qubits inside a superstabilizer,
we can add more qubits to the cat state and can allow every cat state qubit to stabilize a data qubit.
This improvement can be applied with code distance $8$ or higher.
This process is the same as the previous one, except that only one step is required for propagation.
The first SWAP gates overlap with the measurements, then we add
$1$ step for the second SWAP gates,
  $1$ steps for syndrome propagation of ranged pairs,
  $1$ step for a Hadamard gate for $X$ superstabilizer,
  $1$ step for measurements.
to replace a corner cat state qubit with one made inside the superstabilizer, followed by error syndrome propagation and measurement.
In total, $4d+9$ steps are required.

\section{Errors}
\begin{figure}[t]
   \begin{center}
     \includegraphics[width=8cm]{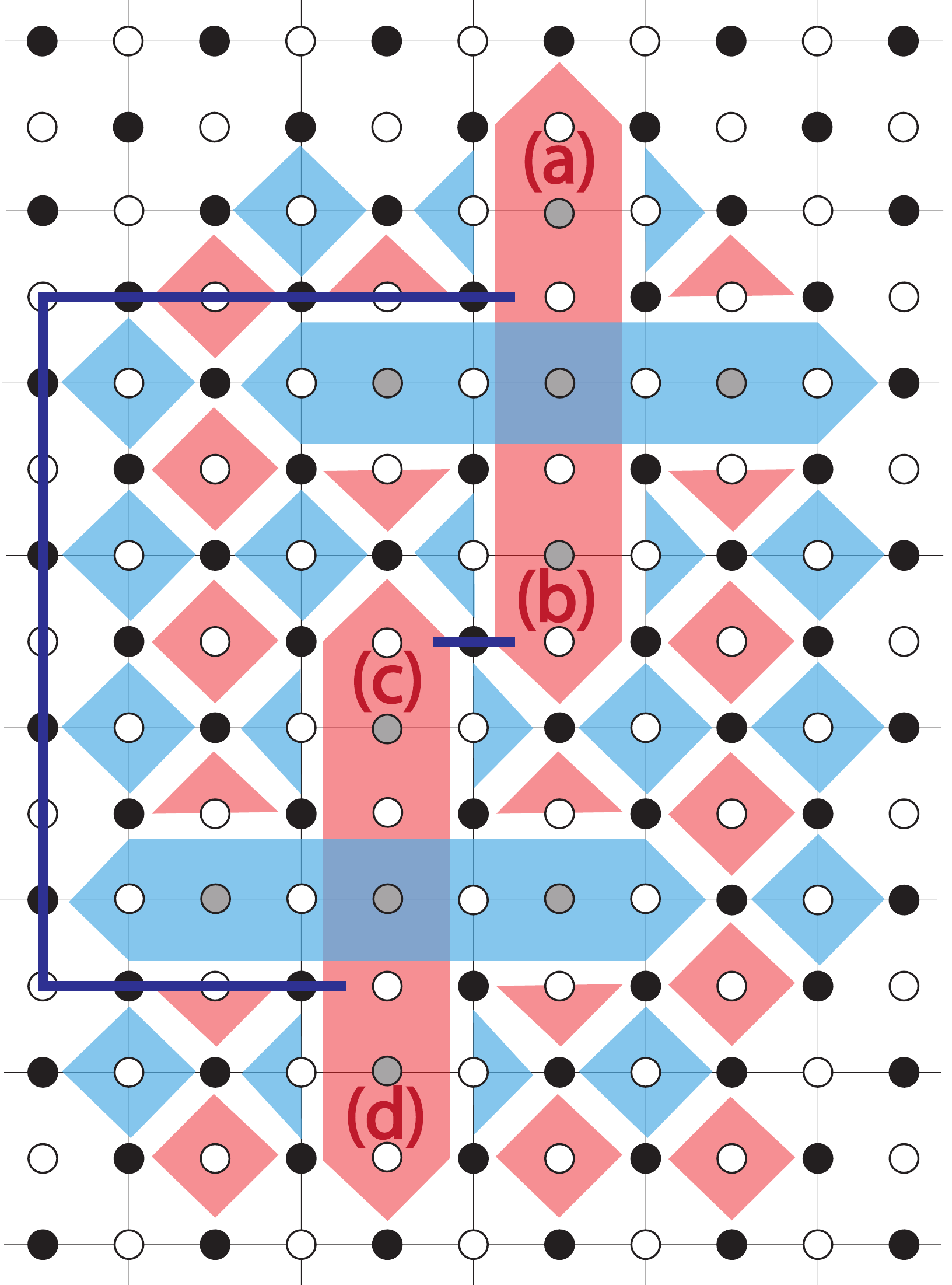}
    \caption{Errors on deformation-based qubits due to the long execution time to measure superstabilizers.
    Either (a) or (b) is a half of a $Z$ superstabilizer.
    A physical $X$ error chain connecting those halves results in a logical $X$ error for this deformation-based qubit.
    So are (c) and (d).
    The set of blue lines describes the product of logical $X$ operators of the two deformation-based qubits.
    }
    \label{fig:operator6}
   \end{center}
  \end{figure}

  Though it might be thought that
  the deep circuit of the superstabilizer measurement results in a
  higher logical error rate than another surface code in which
  any stabilizer requires 8 steps,
  we argue that the deformation-based surface code will exhibit a similar logical error rate with the conventional surface code.
  Figure~\ref{fig:operator6} shows an example of two deformation-based qubits.
  Obviously, any single logical operator is protected by code distance 5,
  as shown in Figure~\ref{fig:correction}.
  Any single operator is protected by normal stabilizers at every $8$ physical steps.
  Therefore conventional error analysis for surface code can be applied.

  The pair of blue lines in Figure~\ref{fig:operator6} indicates the product of
  the two logical qubits' logical $X$ operators.
  In order for a logical error to arise undetected,
  both error chains must occur.
  The short fragment of the operator product between (b) and (c) may occur easily
  and will be detected only by superstabilizer measurements,
  which are completed at every $4d + 5$ physical steps.
  The long fragment of the operator product between (a) and (d) should occur only rarely,
  because the long fragment is protected by normal stabilizers
  and has a longer length than the code distance.
  Therefore the probability that this product operator happens to be executed by errors is
  strongly suppressed, though (b) and (c) are close and $4 d + 5$ physical steps are required to measure superstabilizers.

\begin{figure}[t]
   \begin{center}
    \includegraphics[width=8cm]{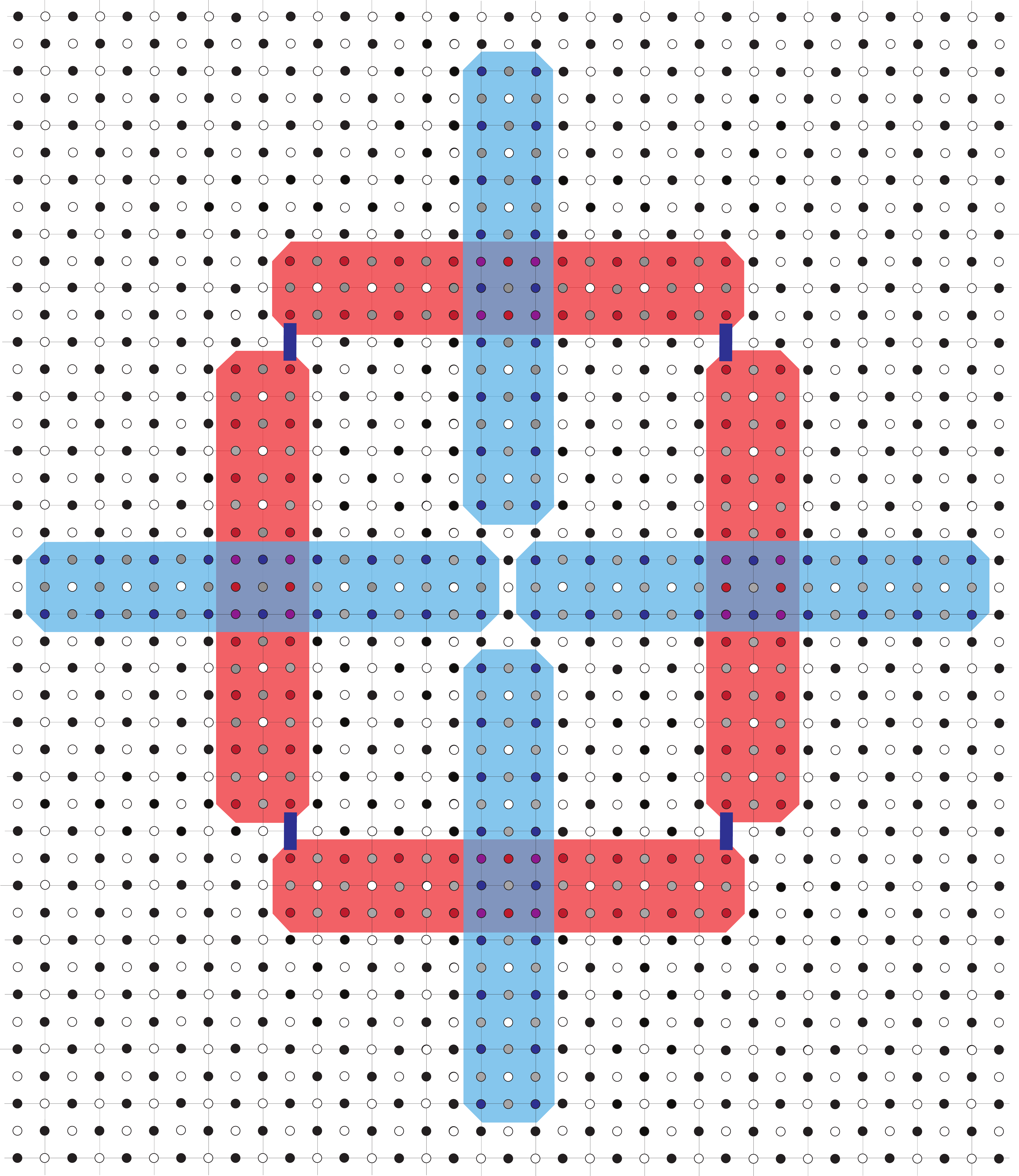}
    \caption{Problematic placement of deformation-based qubits.
    Each deformation-based qubit has code distance 10.
    The shortest combined logical $X$ operator for those four logical qubits
    is only 4, the combination of the shown blue lines.
    }
    \label{problematic_placement}
   \end{center}
  \end{figure}

  Figure~\ref{problematic_placement} shows a problematic placement of deformation-based qubits.
  The code distance of each deformation-based qubit is 10.
  However, the product of the four logical $X$ operators of those deformation-based qubits results in
  the combination of the four blue lines,
  each of which exists between two neighboring $Z$ superstabilizers,
  consisting of only four physical qubits,
  reducing our minimum error chain to $4$.
  Deformation-based qubits must be placed so that their superstabilizers do not form a loop.

\section{Placement design and resource estimation}
\begin{figure}[t]
 \begin{center}
  \includegraphics[width=8cm]{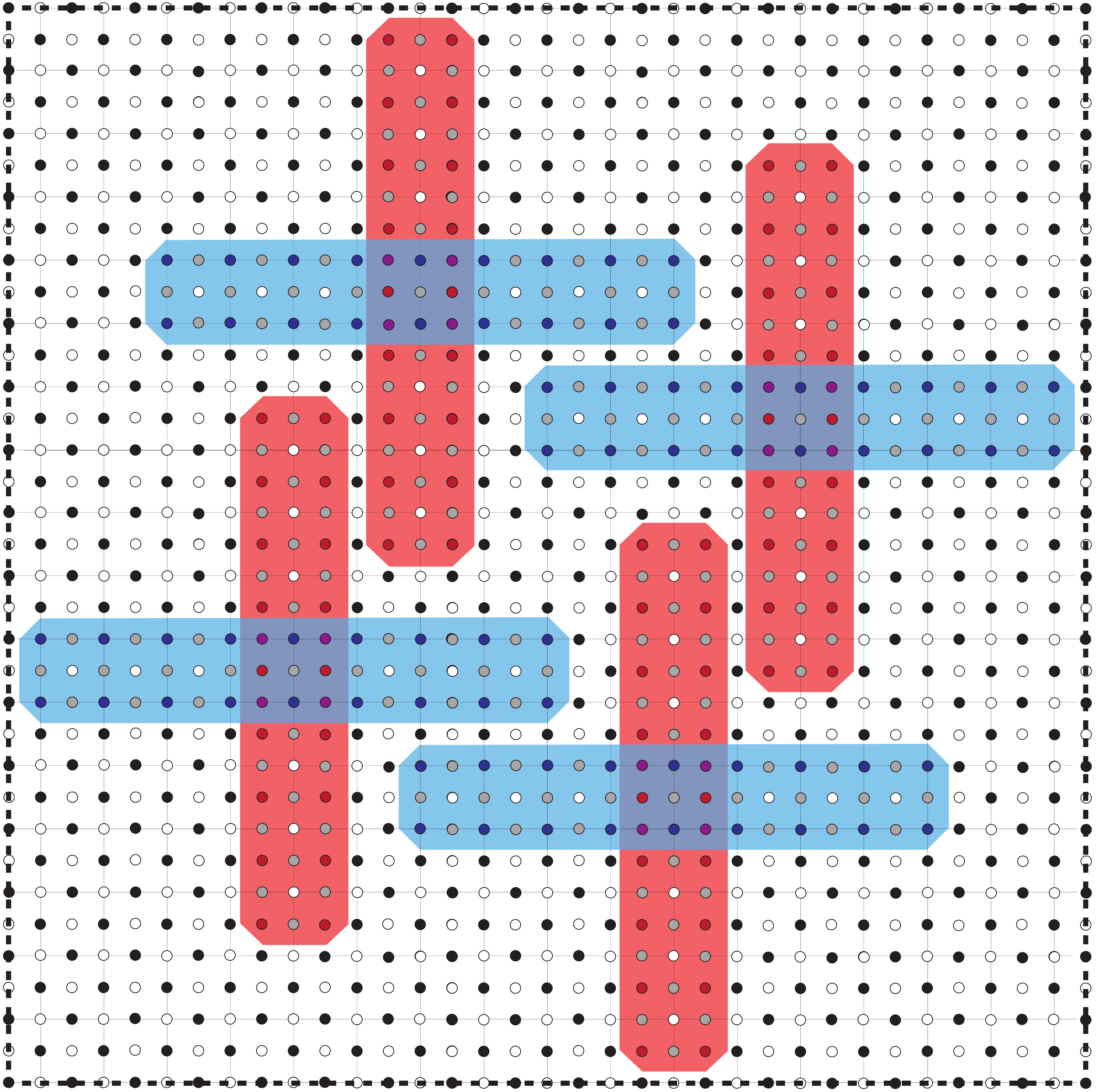}
  \caption{Local placement of the deformation-based surface code.
  There are four logical qubits of distance 10 ($d_o$), however, since the thickness of superstabilizers shorten others' code distance by $1$,
  the reduced code distance $d_s$ is 9.
  This placement enables the four logical qubits to have lattice surgery-like CNOT with other logical qubits.
  For thickness $t=2$, each row and column has $3d_s + 8$ physical qubits
  and $(3d_s+8)^2 = 9d_s^2 + 48d_s + 64$
  physical qubits are required for four logical qubits.
  The dashed box corresponds to the dashed box in Figure \ref{fig:placement_global}.
  }
  \label{fig:placement_local} 
 \end{center}
\end{figure}

\begin{figure*}[t]
 \begin{center}
  \includegraphics[width=15cm]{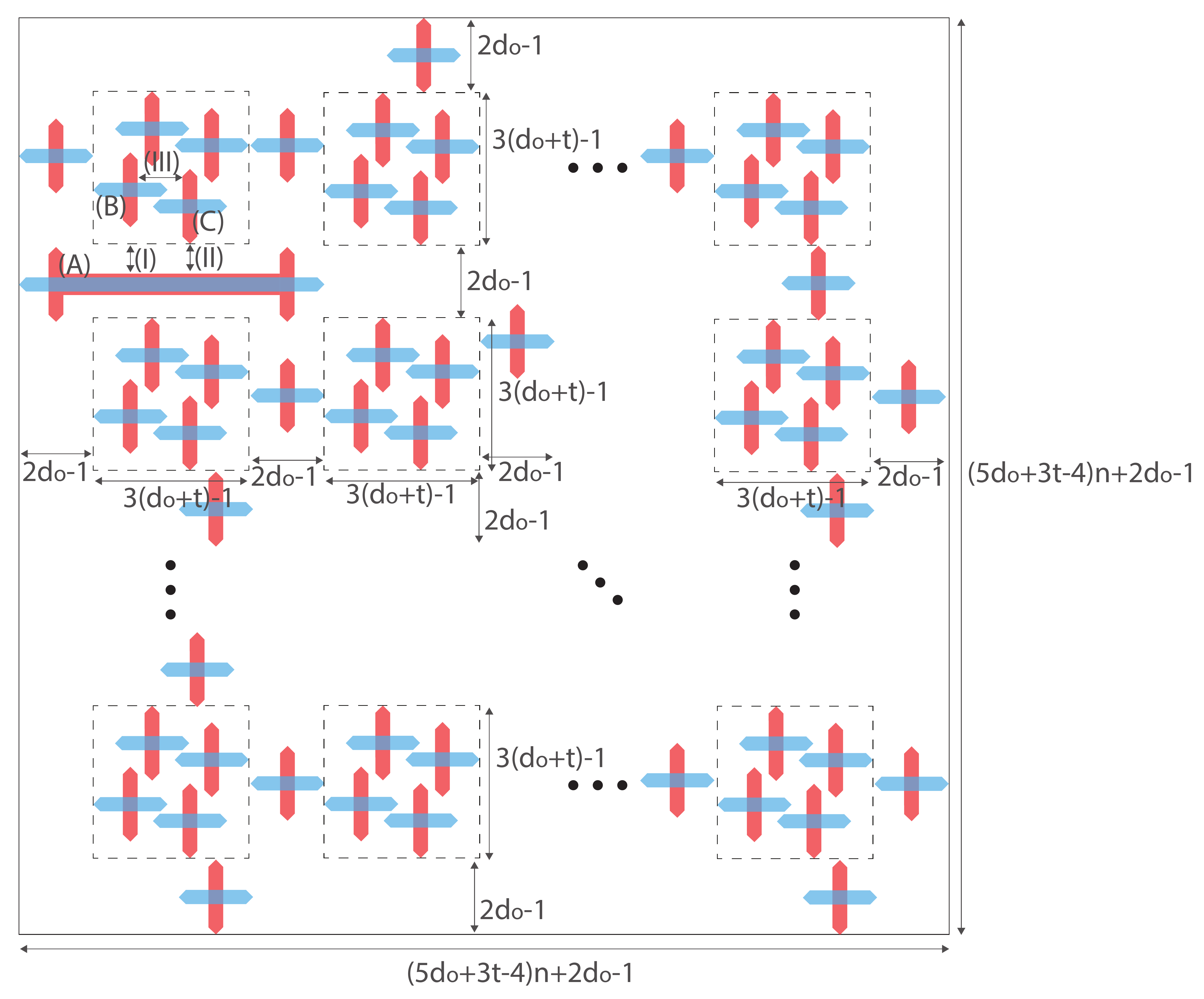}
  \caption{
  Global placement of the deformation-based surface code.
  Each dashed box is the dashed box shown in Figure \ref{fig:placement_local}.
  The spaces between the boxes are paths to move logical qubits and intermediate qubits.
  The deformation-based qubits outside of the dashed boxes are examples of intermediate qubits.
  There are $n$ by $n$ sets of the local placement.
  The lengths includes both data qubits and ancilla qubits, hence
  $2d$ in this figure corresponds to the code distance $d$.
  The stretched qubit indicated with (A) is being routed from location to location.
  To retain the fault-tolerance of (B) and of (C), $(I) + (II)$ needs to be $\frac{d}{2}$ or more,
  therefore (A) is transformed.
  The qubits on the boundary between a local placement set and a path are included both in $3(d_o+t)-1$ and $2d_o-1$,
  hence there are $(5d_o+3t-4)n+2d_o-1$ rows and $(5d_o+3t-4)n+2d_o-1$ columns.
  The total number of physical qubits is $((5d_o+3t-4)n+2d_o-1)^2$,
  for $4n^2$ logical qubits excluding intermediate qubits.
  This placement requires
  $(\frac{5d_o+3t-4}{2})^2 = \frac{25d_o^2 +30d_ot + 9t^2 -40d_o -24t + 16 }{4}$
  physical qubits per logical qubit for enough large $n$.
  }
  \label{fig:placement_global} 
 \end{center}
\end{figure*}

The deformation-based qubits can be placed close together as shown above, however,
there is a restriction that the two $X$ ($Z$) boundaries of a deformation-based qubit cannot be close.
A superstabilizer makes neighbors of stabilizers that were originally distant,
hence shortening distances among stabilizers.
Therefore, we locally set four deformation-based qubits as a box, as shown in Figure~\ref{fig:placement_local},
and globally place the boxes apart to maintain fault-tolerance and to have free space available for routing intermediate qubits
as shown in Figure~\ref{fig:placement_global}.
The code distance of a deformation-based qubit is the shortest number of hops of $Z$ ($X$) stabilizers
between the halves of its $Z$ ($X$) superstabilizer separated by its $X$ ($Z$) superstabilizer.
Each deformation-based qubit in Figure~\ref{fig:placement_local} has code distance $10$
and their superstabilizers' thickness shorten others' code distance by $1$.

Generally, the thickness of a deformation-based qubit may shorten the code distance of neighboring deformation-based qubits.
We set the shortened code distance $d_s$ to be $d_o-t+1$
where $d_o$ is the original code distance and $t$ is the thickness.
%
The number of physical qubits in each row and in each column is
$2(\frac{3(d_o+t) - 1}{2}-1)+1 = 3(d_o+t) - 1$ including both data qubits and ancilla qubits
and this local placement design requires $\frac{(3(d_o+t) - 1)^2}{4} = \frac{9d_o^2 + 18d_ot + 9t^2 - 6d_o - 6t + 1}{4}$ 
physical qubits for a logical qubit.
Therefore, for $t=2$, each row/column has $3d_s + 8$ physical qubits and
and $(3d_s+8)^2 = 9d_s^2 + 48d_s + 64$ physical qubits are required for a logical qubit.

The global placement is shown in Figure~\ref{fig:placement_global}.
  The transformed qubit indicated with (A) is being routed.
  (A) has transformation during moving from one crossroads to another.
  %
  %
  Since the surface code places data qubits and ancilla qubits alternately,
  $2d$ columns/rows are required to have code distance $d$.
  To avoid the situation shown in Figure \ref{problematic_placement},
  $(I)+(II)+(III) \ge 2d$ must be satisfied to guarantee code distance $d$ of (B) and (C).
  Since (III) is $d$, $(I) + (II)$ needs to be $d$ or more hence each of $(I)$ and $(II)$ must be $\frac{d}{2}$ or more.
  Therefore (A) is transformed.
  
  This placement design requires
  $(\frac{5d_o+3t-4}{2})^2 = \frac{25d_o^2 +30d_ot + 9t^2 -40d_o -24t + 16 }{4}$
  physical qubits per logical qubit for enough large $n$.
  Choosing $t=2$,
  $(\frac{5d_s+7}{2})^2 = \frac{25d_s^2+70d_s+49}{4}$
  physical qubits are required for a logical qubit,
  including ancilla qubits.

However, a deformation-based qubit of $t = 2$ in this local placement has a weakness: the superstabilizer itself now appears in a minimal error chain of the code distance. To retain equivalent error suppression strength, we lengthen the crossed superstabilizers one lattice cell, giving an error path of $d$ 4-qubit stabilizers plus the larger, more error-prone superstabilizer.  
As a result, $(\frac{5d_e+12}{2})^2 = \frac{25d_e^2+120d_e+144}{4}$ physical qubits are required for a logical qubit.
  
\begin{figure}[t]
 \begin{center}
  \includegraphics[width=8cm]{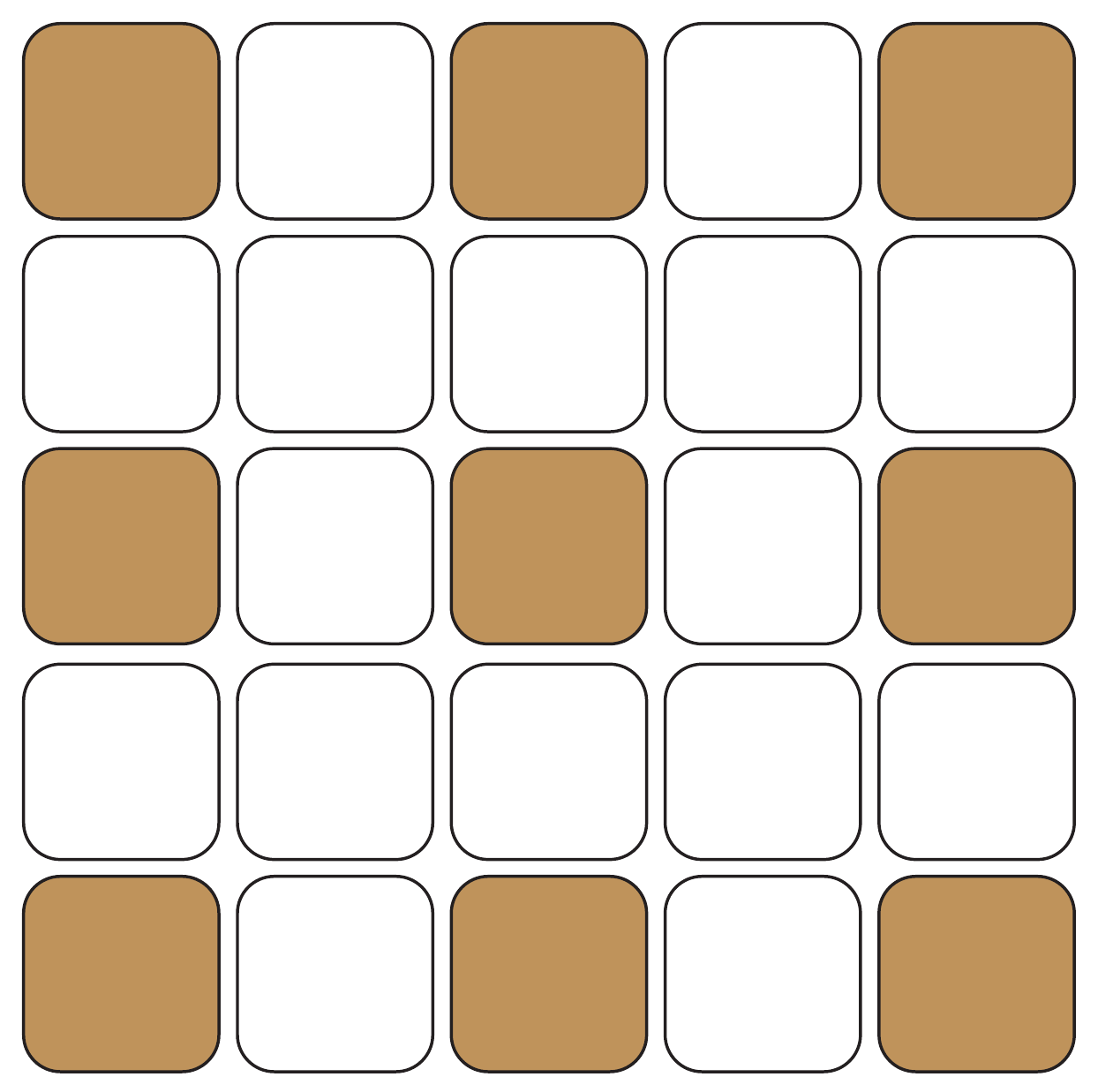}
  \caption{Planar code placement for comparison, after Figure 12 in ~\cite{Horsman:2012lattice_surgery}.
  Each shaded area holds a logical data qubit and
  blank areas are available for intermediate qubits for CNOT gate by lattice surgery.
  Each area has $2d-1$ by $2d-1$ physical qubits, including ancillae.
  }
  \label{fig:planar}
 \end{center}
\end{figure}

  In contrast, the planar code's placement for lattice surgery-based operation,
  shown in Figure \ref{fig:planar}, requires $(4d-2)^2 = 16d^2 -16d + 4$ physical qubits per logical qubit.
  As a result the deformation-based surface code requires $55\%$ fewer physical qubits as the planar code.
  Horsman et al. showed that 
  the number of required qubits for the defect-based surface code is similar to that of the planar code
  in large scale quantum computation,
  so deformation-based surface code also requires fewer physical qubits than the defect-based surface code~\cite{Horsman:2012lattice_surgery}.

We employed the thickness $t=2$ in this example for simplicity.
Using thickness $t=3$ instead 
will shorten the columns and the rows
of a deformation-based qubit.
Because an even code distance has the same error suppression capability as the odd distance just below it,  a $t=2$ logical qubit and a $t=3$ logical qubit should have $2d + 1$ or $2d - 1$ columns/rows, respectively. This allows us to slightly narrow the inter-block channels in Figure~\ref{fig:placement_global}.

\section{Discussion}
We have shown the acceptability of close placement of the deformation-based surface code
by measuring superstabilizers which produce deformation-based qubits;
direct conversion from the defect-based surface code to
the deformation-based surface code, which can be used as state injection
for the deformation-based surface code;
and a lattice surgery-like CNOT gate for the deformation-based qubits
which requires fewer physical qubits than the braiding CNOT gate.
The acceptability of close placement and the space-saving CNOT gate
allow deformation-based qubits
to be packed more tightly than planar code qubits and defect-based qubits.

We have shown theoretical basic concepts but
have not calculated the error suppression ability
since that of the surface code has been investigated well.
The superstabilizers which compose deformation-based qubits
require $4d_e + 9$ steps for stabilizer measurements
where $d_e$ is the effective code distance.
Our placement design preserves logical qubits 
as any logical operator passes through a chain of normal stabilizers that compose part of the code distance, $d_s - 1$.
Hence, by adding $1$ to the code distance,
the long stabilizer measurement does not degrade the error suppression efficiency.
The deformation-based surface code should have residual error rate similar to the conventional surface code of code distance one greater,
and hence conventional error analysis for the surface code can be applied to the deformation-based surface code.

Our design requires
$\frac{25d_e^2+120d_e+144}{4}$
physical qubits for a logical qubit,
compared to the $16d^2 -16d + 4$ physical qubits required in the conventional design.
Our design would halve the resource required to build a large scale quantum computer.

\section*{acknowledgements}
This work is supported by JSPS KAKENHI Grant Number 25:4103 and Kiban B 16H02812.

\bibliography{main}

\newpage
\appendix

\end{document}